\documentclass[11pt]{article}
\pdfoutput=1
\usepackage{epsfig,amscd,amssymb}
\usepackage{amsmath,graphicx}
\newcommand{\comment}[1]{}

\newcommand{\be}{\begin{equation}}
\newcommand{\ee}{\end{equation}}
\newcommand{\bea}{\begin{eqnarray}}
\newcommand{\eea}{\end{eqnarray}}

\newcommand{\zn}{{{\mathbb Z}_n}}
\newcommand{\zth}{{{\mathbb Z}_3}}
\newcommand{\ztwo}{{{\mathbb Z}_2}}

\newcommand{\bb}[2]{\frac{\beta_{#1}}{\beta_{#2}}}

\newcommand{\hhh}[1]{{H^{(#1)}}}
\newcommand{\hh}[2]{{H^{(#1)}_{#2}}}
\newcommand{\HH}{{\cal H}}
\newcommand{\QQ}[1]{\overline{Q}_{#1}}

\newcommand{\ww}{{\rm w}}
\newcommand{\umw}{\phi_{\ww}^{(m)}}

\newcommand{\ub}[2]{\phi^{(#1)}_{#2}}

\newcommand{\vv}[1]{{\eta}^{(#1)}}
\newcommand{\vz}{\vv{0}}

\begin{document}

\title{Free parafermions}
\author{Paul Fendley
\medskip \\ 
Department of Physics, University of Virginia,
Charlottesville, VA 22904-4714 USA\\
}

\smallskip 

\date{November 18, 2013} 

\maketitle

\begin{abstract} 

The spectrum of the quantum Ising chain can be found by expressing the spins in terms of free fermions. An analogous transformation exists for clock chains with ${\mathbb Z}_n$ symmetry, but is of less use because the resulting parafermionic operators remain interacting. Nonetheless, Baxter showed that a certain non-hermitian (but $PT$-symmetric) clock Hamiltonian is ``free'', in the sense that the entire spectrum is found in terms of independent energy levels, with the striking feature that there are $n$ possibilities for occupying each level.  Here I show this directly explicitly finding shift operators obeying a $\zn$ generalization of the Clifford algebra.  I also find higher Hamiltonians that commute with Baxter's and prove their spectrum comes from the same set of energy levels. This thus provides an explicit notion of a ``free parafermion".  A byproduct is an elegant method for the solution of the Ising/Kitaev chain with spatially varying couplings.
\end{abstract} 

\section{Introduction}

A fundamental concept in theoretical physics is that of a free fermion. It not only provides the starting point for perturbative calculations throughout condensed-matter and particle physics, but provides physical information directly via for example band theory. It also plays a fundamental role in many areas of mathematics. For example, the algebra of free-fermionic operators, known in the mathematical literature as a {\em Clifford algebra}, plays a fundamental role in algebraic topology and geometry. This mathematical structure has had an important recent application to physics in the classification of different band theories via their topological invariants \cite{KitK,Ryu}.

Free fermions also play an essential role in a number of important problems seemingly having nothing whatsoever to do with fermions. A famous example is the two-dimensional classical Ising model. Onsager's exact computation of the free energy \cite{Onsager1} is a consequence of the fact that the transfer matrix can be written in terms of bilinears of fermionic operators \cite{Kaufman}. In the quantum spin chain Hamiltonian obtained by taking the anisotropic limit of the classical model, this transformation is widely known as a Jordan-Wigner transformation. The transformation to fermionic variables led to many 
properties being able to be computed exactly, including the spectrum of the quantum Ising spin chain \cite{SML,McCoyWu}. 

This story has now moved in an interesting new direction. Kitaev showed that if fermions instead of spins were taken as the physical degrees of freedom in the quantum Ising chain, the system potentially could  be realized experimentally in a quantum wire placed near a superconductor. Moreover, such a system (now often called the ``Kitaev chain'') has a phase with zero-energy fermionic edge modes and topological order, robust under changes of couplings \cite{KitMajorana}. This result has prompted a huge amount of activity, because of potential applications to topological quantum computation \cite{Jason}. There is even a 2+1-dimensional spin system that can be mapped onto free fermions coupled to a background gauge field \cite{Kithoneycomb}.

Because of these spectacular successes, it is natural to look for generalizations. An infinity of conserved charges makes exact computations possible in integrable models \cite{Baxbook}, but these methods are quite different from the free-fermion methods, and do not yet yield results like the complete spectrum. Systems solvable by free-fermion methods remain very special. 

There are however many indications that certain models with $\zn$ symmetry have similar structure to free-fermionic models. In these models, the Ising two-state systems are replaced by $n$-state ``clock'' variables. The Fradkin-Kadanoff transformation to parafermionic operators \cite{FK} is the clock-chain analog of Jordan-Wigner transformation to fermions in Ising. Derivations of the results for the $\zn$-invariant system are considerably more involved than those for Ising/Kitaev chain, because parafermionic operators pick up a phase $e^{2\pi i/n}\equiv\omega$ when their order is exchanged, and this seemingly innocuous generalization makes the model interacting. Even in integrable clock Hamiltonians \cite{FZ1,Gehlen,Baxterclock}, correlators in the corresponding parafermionic conformal field theory \cite{FZ2,RR} do not obey Wick's theorem. It was even boldly stated in \cite{para} that ``there is no such thing as a free parafermion in 1d''. 
Nevertheless, the integrable chiral Potts model \cite{Gehlen,Perkoverview} has many relations with free-fermionic models, including for ``superintegrable'' couplings, the same algebraic structure as Onsager found in the Ising model \cite{Onsager1,Dolan,Gehlen,Davies}. Moreover, the chiral clock chain has a phase with robust topological order and an exact parafermionic edge zero mode \cite{para}.

The most striking connection comes from a  particular (non-hermitian) clock Hamiltonian. Baxter showed  \cite{Baxterclock} that the spectrum is very much a $\zn$ analog of a free-fermion spectrum. Subsequently, a two-dimensional classical model related to this clock chain was found, and a host of fascinating relations with the integrable chiral Potts model were derived \cite{BS,Baxrecent}. 
The derivation of this result is somewhat indirect, relying on results arising from the integrable chiral Potts model \cite{Gehlen,Perkoverview}. 
The purpose of this paper is to rederive the spectrum of Baxter's clock Hamiltonian directly, and so give a precise notion of what a ``free parafermion'' is.

To be more explicit: the basic property of a quantum free-fermionic system is that all the energies in the spectrum are all given by
\begin{equation}
E=\pm \epsilon_1 \pm \epsilon_2 \pm \dots \pm \epsilon_N\ .
\label{ffE}
\ee
Each $\epsilon_k$ is called an energy {\em level}, and 
all the eigenvalues of the Hamiltonian are specified by choosing each $\pm$ independently. The key property that makes the system free is that the values of the $\epsilon_k$ are the same for all $2^N$ states of the theory; they do not vary with the choice of the $\pm$ signs. In the Ising/Kitaev chain on $L$ sites,  the values of the $\epsilon_k$ are determined by diagonalizing a $L \times L$ matrix, not the $2^L$-dimensional Hamiltonian. Despite the use of $k$ to label the energy levels, this result does not require translation invariance or periodic boundary conditions or spatially uniform couplings or even short-range interactions. Changing any of these modifies entries in the matrix, but the spectrum remains of the same form as long as the Hamiltonian is bilinear in the fermions.

\begin{figure}[t]
 \begin{center}
 \includegraphics[scale=0.45]{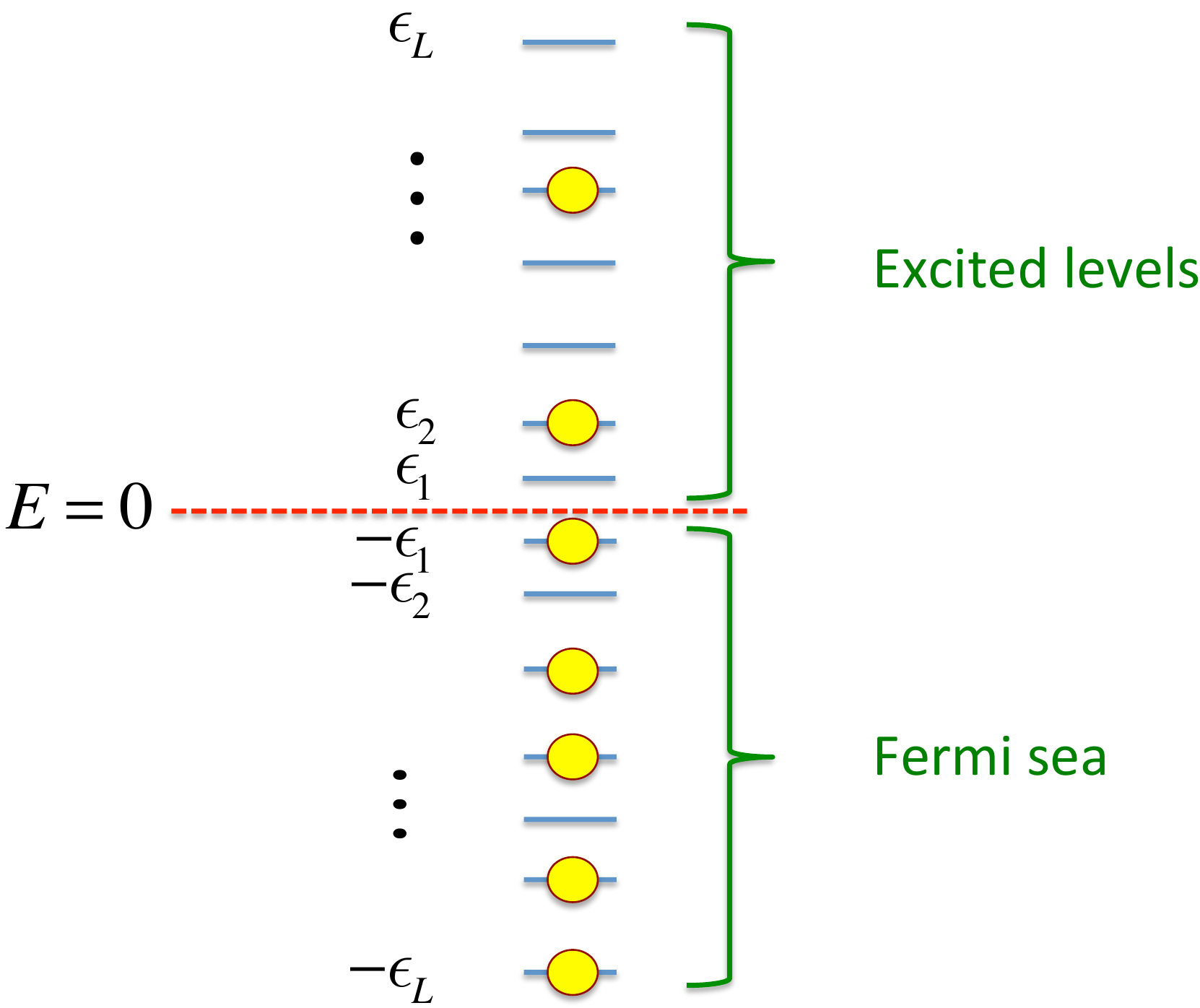}
\end{center}
 \caption{How levels are filled for free-fermion theories}
 \label{fig:fermionlevels}
\end{figure}

A convenient way of visualizing this free-fermi spectrum is given in figure \ref{fig:fermionlevels}. Each level corresponds to two states in the picture, one in the Fermi sea. Choosing a $-$ sign corresponds to filling a state in the Fermi sea, while choosing a $+$ sign corresponds to filling an excited state. Thus each energy  is given by making this choice for each level, yielding all $2^L$ energies in the spectrum.  The ground state corresponds to filling the entire Fermi sea, i.e.\ choosing all minus signs.

Baxter's result for the spectrum of his non-hermitian clock Hamiltonian (given in (\ref{Hclock}) below) on an $L$-site chain with free boundary conditions is very similar \cite{Baxterclock}:
\be
E=\omega^{p_1} \epsilon_1 +\omega^{p_2} \epsilon_2 +\dots + \omega^{p_L} \epsilon_L
\label{spectrum}
\ee
for any choice of the $p_k=0\dots n-1$. This gives all $n^L$ eigenvalues in the spectrum, with
the $\epsilon_k$ coming from the eigenvalues of a $L\times L$ matrix just as in the free-fermion case. This matrix is independent of how the $p_k$ are chosen. Even more strikingly, with an appropriate parametrization of couplings, the matrix is independent of $n$ and so identical to that of the Ising/Kitaev chain.

Since the values of all the $\epsilon_k$ are independent of the choice of $p_k$, it is natural to call this model ``free''. The natural followup question is then ``free what?".  Since the energies are complex for $n>2$, there is no notion of a ground state or quasiparticles above it. Nonetheless, it is still natural to call the $\epsilon_k$ energy levels, and to think of the spectrum (\ref{spectrum}) as coming from quasiparticles occupying these levels. One of the many interesting features of this model though is that the notion of ``filling'' requires substantial modification. Each of the $L$ levels is occupied with one quasiparticle, but there are $n$ possible choices for each $k$, instead of just ``empty" or ``filled" as in the free-fermion case.  This idea is illustrated in figure \ref{fig:paralevels}, and provides a very nice example of what Haldane termed ``exclusion statistics'' \cite{Haldane}. For a lack of a better name, I call these quasiparticles ``free parafermions''.

\begin{figure}[ht]
 \begin{center}
 \includegraphics[scale=0.4]{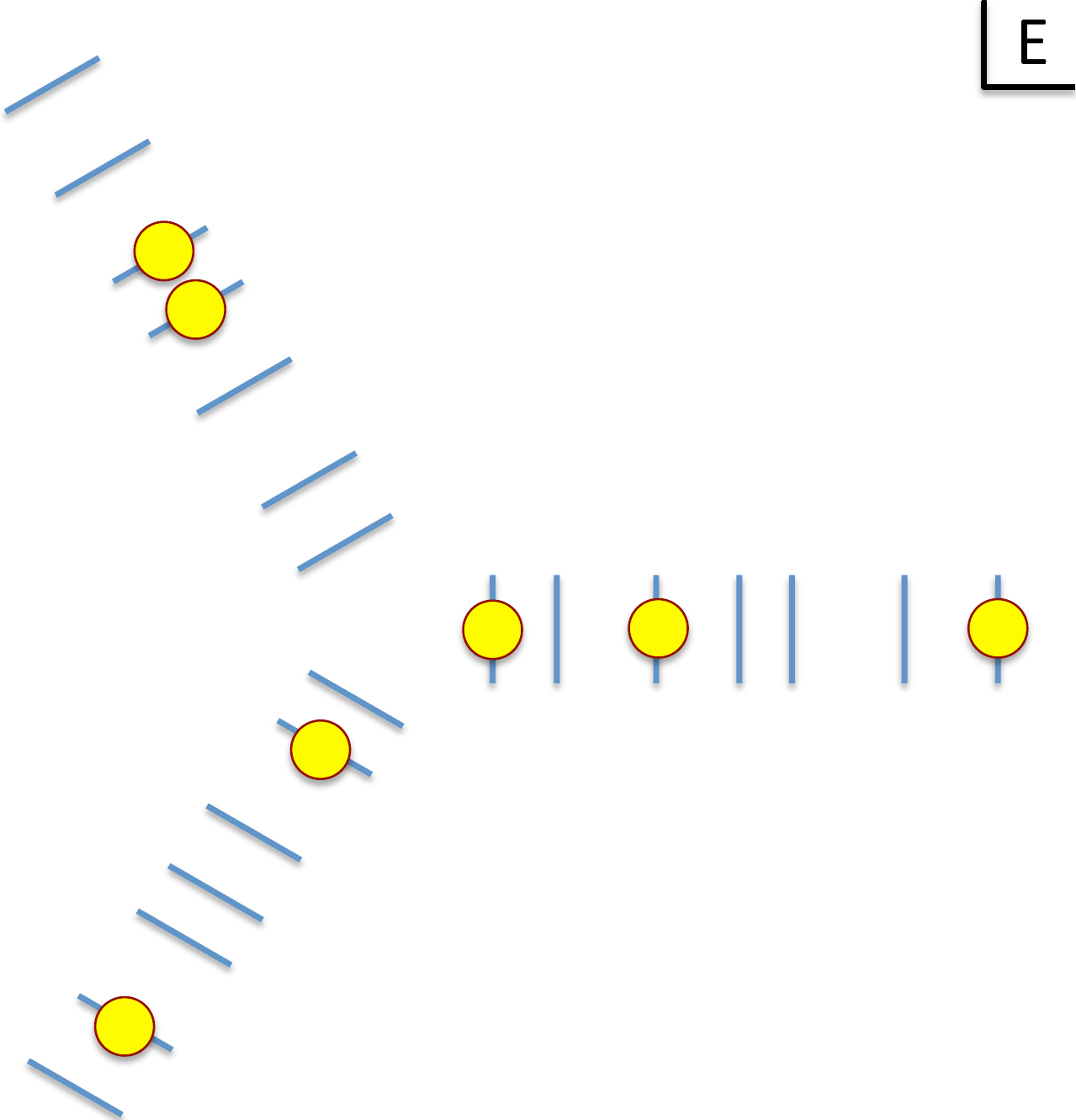}
\end{center}
 \caption{Filling levels in the $\zth$ free-parafermion theory}
 \label{fig:paralevels}
\end{figure}

This paper provides a detailed account of where this beautiful structure comes from, and generalizes it as well. I derive both non-local conserved currents and ``higher'' Hamiltonians $\hhh{m}$ that commute with Baxter's and so can be diagonalized simultaneously. The higher Hamiltonians are analogous to those formed in the fermion case from pairs of fermions not nearest-neighbor, but the expressions here are considerably more complicated. The energy spectrum of each of these Hamiltonians is 
\be
E^{(m)}=(\omega^{p_1} \epsilon_1)^m +(\omega^{p_2} \epsilon_2)^m +\dots + (\omega^{p_L} \epsilon_L)^m\ .
\label{higherspectrum}
\ee
where the $\epsilon_k$ are those of Baxter's Hamiltonian, and the $p_k$ are chosen in the same way.

The plan of this paper is as follows. In section \ref{sec:Ising} I give a very general solution of the Ising/Kitaev chain for spatially varying couplings and open boundary conditions. The method used is a variation of that used in \cite{McCoyrandom} for the Ising transfer matrix with random couplings in one direction. Some interesting polynomials arise in this analysis, making it possible to show that the Hamiltonian can be written as the sum of commuting (albeit non-local) projection operators.

In section \ref{sec:zn}, I introduce Baxter's clock Hamiltonian and show that it has both non-local and local conserved quantities, the latter the higher Hamiltonians. The conserved quantities derived in section \ref{sec:zn} do not require that $\omega$ be a root of unity, but in section \ref{sec:Hwn} I show how some remarkable simplifications occur when $\omega^n=1$. While the model is in some ways much more complicated than the free fermion case $n=2$, it has many features in common. In particular, this enables the derivation in section \ref{sec:shift} of ``shift'' operators that, when acting on an eigenstate of the Hamiltonian(s), shift between different values of $p_k$ for a given level. In section \ref{sec:spectrum} I use the higher Hamiltonians and the shift operators to derive the spectrum (\ref{higherspectrum}) for all $m$, reproducing Baxter's result for $m=1$. From this one infers a generalization of the Clifford algebra useful for solving models for all $n$.

Since the energies in (\ref{higherspectrum}) are not real, physical applications may seem to be limited. Nonetheless, I believe that these results have not only formal significance but some more general potential applications as well. These, as well as interesting formal directions to pursue, are discussed in section \ref{sec:conclusion}, the conclusion. Some fairly gruesome technical results are collected in the appendices.

\section{The quantum Ising chain}

\label{sec:Ising}

The quantum Ising chain is the $n=2$ case of the clock models discussed in this paper. That its exact spectrum can be computed using free fermions is a fundamental result of theoretical physics \cite{Kaufman}. The computation is quite simple and elegant \cite{SML}, despite an undeserved reputation to the contrary. I compute the spectrum here following a slightly different path than customary, first deriving the raising/lowering operators. This method is technically simpler than the usual ``Bogoliubov'' transformation of the Hamiltonian. It also is applicable to the case of spatially varying couplings and/or open boundary conditions; the customary method exploits translation invariance and so requires periodic boundary conditions and uniform couplings.

\subsection{The quantum Ising chain in terms of fermions}
\label{sec:Isingfermions}

The Hilbert space for the quantum Ising chain is $({\mathbb C}^2)^{\otimes L}$, i.e.\ it consists of a two-state quantum system, i.e.\ a ``spin'' 1/2 particle, at each of the $L$ sites. 
The Hamiltonian is comprised of two types of terms, those that flip a spin at a given site, and those that describe an interaction energy between adjacent spins. Precisely, with open boundary conditions the Hamiltonian is
\be
H_{\rm IM} = -  \sum_{j=1}^L t_{2j-1}\sigma^x_j - \sum_{j=1}^{L-1}t_{2j} \sigma^z_j\sigma^z_{j+1}
\label{HIsing}
\ee
where the operators $\sigma^a_j = 1\otimes \dots 1\otimes  \sigma^a\otimes 1\dots \otimes 1$ act non-trivially with a Pauli matrix on the two-state system at site $j$. The $t_a$ are arbitrary real couplings; typically the $t_{2j-1}$ are fixed to be one value $t$ and the $t_{2j}$ are fixed to be $J$, with the critical point occurring at $t=J$. However, it is only marginally more difficult to find the spectrum for spatially varying couplings, and so I will do so here. This Hamiltonian has a ${\mathbb Z}_2$ symmetry under flipping all the spins. For reasons to be apparent shortly, it is natural to name this spin-flip operator $(-1)^F$, so that
\be
(-1)^F = \prod_{j=1}^L \sigma^x_j \ .
\label{min1F}
\ee
This operator indeed squares to $1$, and it is easy to check that 
$[(-1)^F,H_{\rm IM}] = 0$. 

Anticommuting fermionic operators are defined by a non-local combination of the spin operators known as a Jordan-Wigner transformation \cite{SML}. Precisely, 
for each site of the chain, there are two ``Majorana'' fermion operators defined by
\begin{eqnarray}
\psi_{2j-1} = \left(\prod_{k=1}^{j-1} \sigma^x_k\right) \sigma^z_j \ ,\qquad
\psi_{2j} =-i\psi_{2j-1}\sigma^x_j= i \left(\prod_{k=1}^{j}\sigma^x_k\right)  \sigma^z_j\ .
\label{bdef}
\end{eqnarray}
Each operator $\psi_j$ is hermitian and squares to 1. 
Because of the non-local ``strings'' attached, operators at different points no longer commute. Instead, they are fermionic and so anticommute: 
\begin{equation}
\{ \psi_a, \psi_b\} = 
2\delta_{ab}\ .
\label{Cliffa}
\ee
for all $a,b=1\dots 2L$. Such an algebra is known as a {\em Clifford algebra}. Clifford algebras have a variety of fascinating properties, several of which are discussed below.
The open-chain Hamiltonian (\ref{HIsing}) rewritten in terms of these fermionic operators is 
\be
H_{\rm IM} = i \sum_{a=1}^{2L-1} t_a\,\psi_{a+1}\psi_{a} \ .
\label{HIF}
\ee
A convenient feature of the Majorana fermionic basis apparent here is that it puts the flip and potential terms on an equal footing. The fact that $H_{\rm IM}$ is a sum of fermion bilinears makes possible the computation of the spectrum described below. 

The Ising Hamiltonian can be rewritten in terms of the standard complex fermions $c^{\dagger}_j= (\psi_{2j-1}+i\psi_{2j})/\sqrt{2}$. It then includes ``Cooper-pairing'' terms involving $c_j c_{j+1}$ and $c^\dagger_j c^\dagger_{j+1}$, and so conserves only fermion number mod 2. The corresponding symmetry generator $(-1)^F$ commutes with any product of an even number of fermion operators, while it anticommutes with any product of an odd number. It thus is the product of all the Majorana fermion operators:
$(-1)^F = i^{L}\prod_{a=1}^{2L} \psi_a\ ,$
which indeed is the spin-flip $\ztwo$ symmetry generator from (\ref{min1F}).

\subsection{Computing the spectrum from raising and lowering operators}
\label{sec:Isingshift}

I explain here how to construct raising and lowering operators in the quantum Ising chain. A fermionic raising or lowering operator $\Psi$ has the property that
 \be
[H, \Psi] = 2\epsilon \Psi\ ,
\label{HPsicomm}
\ee
so that acting with $\Psi$ on an eigenstate of $H_{\rm IM}$ either annihilates it or gives another eigenstate with energy shifted by $2\epsilon$. The spectrum then is found by exploiting the fact that the raising and lowering operators obey a Clifford algebra. 

The key to the computation is the fact that a commutator of any bilinear in fermions obeying the Clifford algebra (\ref{Cliffa}) with anything linear yields something linear:
$$[\psi_{a}\psi_b,\psi_c] = \psi_b\delta_{ac} - \psi_a\delta_{bc}$$
for any $a$, $b$ and $c$. Let 
$$\Psi=\sum_b i^b\mu_b \psi_b$$ be such a linear combination, with this and all sums in this section over $a$ or $b$ running from $1$ to $2L$ unless labeled otherwise. Commuting this with $H_{\rm IM}$  then gives
\be
\Psi'= [H_{\rm IM}, \Psi]=\sum_a \mu_a' \psi_a,\qquad\hbox{ with }\quad \mu'_a = 2(t_a\mu_{a+1} + t_{a-1}\mu_{a-1})
\label{psiprime}
\ee
where $t_0=t_{2L}=\mu_0=\mu_{2L+1}\equiv 0$. This can conveniently be written in matrix form as 
$$\begin{pmatrix}\mu_1'\\ \mu_2'\\ \vdots\\  \\ \mu_{2L}' \end{pmatrix} =
2\begin{pmatrix}
0&t_1&0&\dots\\
t_1&0&t_2&  \\
0&t_2&0&\\
\vdots&&&&t_{2L-1}\\
&&&\ t_{2L-1}&0
\end{pmatrix}
\begin{pmatrix}\mu_1\\ \mu_2\\ \vdots\\  \\ \mu_{2L} \end{pmatrix} 
$$
i.e.\  
\be
\mu'_a= 2\sum_{b} {\cal F}_{ab} \mu_b\quad \hbox{ with }\quad {\cal F}_{ab} = t_a(\delta_{a,b+1}-\delta_{b,a+1})\ .
\label{calFdef}
\ee
The hermitian $2L\times 2L$ matrix ${\cal F}$ thus describes the linear action of commuting $H_{\rm IM}$ with anything linear in the fermions. It has determinant $(-1)^L\prod_{j=1}^L t_{2j-1}^2$.

The raising and lowering operators follow immediately from the eigenvectors of ${\cal F}$. Namely, if the $\mu_a$ form an eigenvector of ${\cal F}$ with eigenvalue $\epsilon$, then the corresponding $\Psi$ satisfies (\ref{HPsicomm}).  Eigenvalues of ${\cal F}$ form pairs of opposite sign, as is easily checked by sending $\mu_b\to (-1)^b\mu_b$. Thus defining $\Psi_{+k}$ to have eigenvalue $\epsilon_k>0$, then $\Psi_{-k}\equiv \Psi^\dagger_{+k}$ has eigenvalue $\epsilon_k$.
The eigenvectors are thus conveniently labeled by index $\pm k$ with $k=1\dots L$ so that
\be
[H,\,\Psi_{\pm k}]=\pm 2\epsilon_k \Psi_{\pm k} 
\label{Iraising}
\ee
with the convention $\epsilon_k>0$; all $t_a$ are assumed to be non-zero so that all eigenvalues of ${\cal F}$ are non-vanishing. 
If boundary conditions were periodic and the couplings uniform so that the system were translation invariant, $k$ would also label momentum. However, it is not necessary to have translation invariance to describe the spectrum in terms of free fermions.

The $\Psi_{+k}$ are raising operators; acting on an eigenstate of $H_{\rm IM}$ with energy eigenvalue $E$ each either annihilates it or gives an eigenstate with energy $E+2\epsilon_k$. Likewise, $\Psi_{-k}$ is a lowering operator that either annihilates an eigenstate or gives another with energy $E-2\epsilon_k$. The existence of these operators is consistent with the free-fermion energy spectrum (\ref{ffE})
$$
E=\pm \epsilon_1 \pm \epsilon_2 \pm \dots \pm \epsilon_{L}
$$
described in the introduction. Consider a eigenstate of ${\cal F}$ with eigenvalue having some choice of the $\pm$ signs. Then $\Psi_{+k}$ annihilates any eigenstate where the $k$th sign is positive, while if it is negative it acts non-trivially, giving an eigenstate where the $k$th sign is now positive, leaving the others unchanged. The lowering operators do the opposite. Because $\{(-1)^F,\Psi_{\pm k}\}=0$, any $\Psi_{\pm k}$ necessarily maps between sectors with even and odd eigenvalues of $(-1)^F$.

For the quantum Ising chain with $L$ sites, there are $2^L$ states, and so $2^L$ eigenvalues of $H_{\rm IM}$. It is thus natural to expect that each of the energies in (\ref{ffE}) occurs exactly once in the spectrum and that all the eigenstates can be found by acting with the raising operators on the ground state, that with eigenvalue (\ref{ffE}) with all minus signs. The simplest way to  prove that the spectrum is indeed is free-fermion is to find the algebra obeyed by the raising and lowering operators. 
First of all, note that each squares to zero. Letting $\mu_b$ form an eigenvector of ${\cal F}$ gives
$$
\Psi^2=\sum_{a,b}i^{a+b}\mu_a\mu_b \psi_a\psi_b = \sum_b (-1)^b\mu_b^2 = 0\ ;
$$
terms with $a\ne b$ in the sum cancel because then $\psi_a$ and $\psi_b$ anticommute, while as noted above $\mu_b$ and $(-1)^b\mu_b$ correspond to eigenvectors of ${\cal F}$ with opposite eigenvalue and so have vanishing inner product. This means e.g.\ that acting with $\Psi_{+k}$ twice cannot give an eigenvector with eigenvalue shifted by $4\epsilon_k$, a necessary condition for (\ref{ffE}) to apply. More generally, consider the anticommutator of two operators $\Psi$ and $\chi$ built from eigenvectors $\mu_a$ and $\nu_b$:
$$\{\Psi,\,\chi\}= \sum_{a,b}i^{a+b}\mu_a\nu_b\{\psi_a,\,\psi_b\} = 2\sum_a(-1)^a\mu_a\nu_a\ .
$$
By orthonormality of eigenvectors, this vanishes unless $\nu_a=(-1)^a\mu_a$, i.e.\ the anticommutator is between $\Psi_{+k}$ and $\Psi_{-k}$. These anticommutation relations are therefore 
\begin{equation}
\{\Psi_{\pm k},\Psi_{\pm k'}\} = 0\ ,\qquad \{\Psi_{\pm k},\Psi_{\mp k'}\} = N_k \delta_{kk'}\ ,
\label{Cliffk}
\ee
where $N_k = 2\sum_a |\mu_a|^2$ for the eigenvectors with eigenvalue $\pm 2\epsilon_k$. It follows that
\be
(\Psi_{+k} \pm \Psi_{-k})^2 = \pm N_k
\label{Psisq}
\ee
with $N_k>0$. This means the operators 
$$\frac{1}{\sqrt{N_k}}(\Psi_{+k} + \Psi_{-k})\quad\hbox{and}\quad \frac{i}{\sqrt{N_k}}(\Psi_{+k} - \Psi_{-k})$$
satisfy the same Clifford algebra (\ref{Cliffa}) that the $\psi_a$ do.

These anticommutation relations are enough to prove that the spectrum of $H_{\rm IM}$ must be (\ref{ffE}). Namely, the relation (\ref{Psisq})  shows that $(\Psi_{+k} + \Psi_{-k})^2$ is a non-zero constant, so that no state in the Hilbert space can be annihilated by both $\Psi_{+k}$ and $\Psi_{-k}$. Thus given any eigenstate of $H_{\rm IM}$, there must be another one with energy either $E+2\epsilon_k$ or $E-2\epsilon_k$ for all values of $k=1\dots L$. The only way this can be true for all $2^L$ eigenstates of $H_{\rm IM}$ is if the spectrum is of the form (\ref{ffE}) up to a constant shift. It is easy to see that there is a charge conjugation symmetry showing the spectrum of $H_{\rm IM}$ is invariant under $E\to -E$, so this constant shift must vanish. Moreover, the spectrum can be described in terms of independent energy levels as illustrated in figure \ref{fig:fermionlevels}. 
Because $\Psi_{\pm k}$ and $\Psi_{\pm k'}$ anticommute for $k\ne k'$, raising or lowering the state in a given level $k$ does not affect any other levels. Thus one can imagine the ground state as a sea of fermions with all the negative-energy levels filled. The operator $\Psi_k$ acting on it then creates a ``free'' fermion excitation of energy $2\epsilon_k$.

It is not difficult to generalize the above analysis to the case of periodic boundary conditions. This merely requires that the ``round-the-world'' interaction between spins at sites $L$ and $1$ be bilinear when rewritten in terms of the fermions. This results in other non-zero entries to the matrix ${\cal F}$, but otherwise the preceding analysis is unchanged.
In fact, any bilinear in the fermions can be included in the Hamiltonian; this merely results in still more non-zero entries in ${\cal F}$.

These free-fermionic operators thus reduce the computation of the $2^L$ values in the spectrum of $H_{\rm IM}$ to finding the eigenvalues of a $2L\times 2L$ matrix ${\cal F}$. For all $t_a=t$, these eigenvalues are $\pm 2\epsilon_k = \pm 4t\cos(\pi k/(2L+1))$, so that the spectrum is gapless over the ground state; this is the critical point of the Ising chain. There of course is not a closed-form expression for the eigenvalues of ${\cal F}$ for arbitrary couplings. However, defining
\be
Q_{2L}(u)={\rm det}(u^{1/2}-{\cal F})\ .
\label{Q2def}
\ee
gives a polynomial in $u$ of order $L$. Letting $u_k$ be the roots of this polynomial, then $\epsilon_k=\sqrt{u_k}$. 

An explicit expression for $Q_{2L}$ follows can be found using a recursion relation for this determinant. To find this, it is convenient to define $Q_{2L-1}={\rm det}(u^{1/2}-{\cal F}')$, where ${\cal F}'$ is ${\cal F}$ with the last row and last column deleted. Physically, ${\cal F}'$ arises from a chain with free boundary conditions at $j=1$ as before, but fixed at $j=L$, i.e.\ setting $t_{2L-1}=0$. Then the usual cofactor relation for determinants gives for $a\ge 1$
\be
Q_{a+1}(u) = u^{1/2}Q_{a}(u) - \gamma_{a}\, Q_{a-1}(u) \ 
\label{Qrecur}
\ee
where $\gamma_a\equiv t_a^2$, and $Q_0(u)=1$ and $Q_1(u)=u^{1/2}$.  
Explicitly,
\begin{eqnarray*}
Q_2 &=& u-\gamma_1\ ,\\
Q_3 &=& u^{3/2}- (\gamma_1 + \gamma_2)u^{1/2}\ ,\\
Q_4 &=& u^2- (\gamma_1+\gamma_2+\gamma_3)u+\gamma_1\gamma_3\ ,\\
Q_5 &=& u^{5/2}- (\gamma_1+\gamma_2+\gamma_3+\gamma_4)u^{3/2}+(\gamma_1\gamma_3+\gamma_1\gamma_4+\gamma_2\gamma_4)u^{1/2}\ ,\\
Q_6 &=& u^3 - (\gamma_1+\gamma_2+\gamma_3+\gamma_4+\gamma_5)u^2+(\gamma_1\gamma_3+\gamma_1\gamma_4+\gamma_1\gamma_5+\gamma_2\gamma_4+\gamma_2\gamma_5+\gamma_3\gamma_5)u-\gamma_1\gamma_3\gamma_5.
\end{eqnarray*}
The roots of the $Q_a$ are guaranteed to be real since ${\cal F}$ is hermitian. Looking at the explicit expressions for the $Q_a$ shows there is an ``exclusion rule'' in the polynomial: the products $\gamma_a\gamma_{a+1}$ and $\gamma_a^2$ never occur in the coefficients. This suggests the general expression
\be
Q_a(u)= u^{a/2}\sum_{l=0}^{[a/2]}(-u)^{-l}  
\sum_{b_1=1}^{a-(2l-1)}\,\sum_{b_2=b_1+2}^{a-(2l-3)} \cdots \sum_{b_l=b_{l-1}+2}^{a-1}\  \prod_{j=1}^{l} \gamma_{b_j}\ ,
\label{Qdef}
\ee
where $[r]$ is the integer part of $r$.
This explicit expression is easily proved by showing that it satisfies the recursion relation (\ref{Qrecur}).

It is worth noting that since $Q_1=u^{1/2}$, the recursion relation (\ref{Qrecur}) resembles the tensor product (or ``fusion'') of the representations of $SU(2)$ when all the $t_a=1$, while the truncation $Q_{2L}(u_k)=0$ is reminiscent of what happens for the quantum-group deformation $U_q({sl_2})$ when $q$ is a root of unity. Intriguingly, the more general recursion relation for $\gamma_a\ne 1$ and polynomials also arose in \cite{steve} for counting certain types of closely packed loop configurations on two-dimensional lattices.

\subsection{The Hamiltonian in terms of raising/lowering operators}
\label{sec:bogo}

It is illuminating to rewrite the Hamiltonian in terms of the $\Psi_{\pm k}$. The simplest way of doing this is to note that the operators
\be
P_{\pm k} =\frac{1}{N_k}\Psi_{\pm k}\Psi_{\mp k} 
\label{projop}
\ee
are projection operators, i.e.\ they satisfy $(P_{\pm k})^2=P_{\pm k}$, as is easily verified by using the algebra (\ref{Cliffk}). It also follows that  
$$P_{+k}+P_{-k}=1\ ,\qquad P_{\pm k}\Psi_{\mp k} =\Psi_{\pm k}P_{\pm k}=0\ ,\qquad [P_{\pm k}, P_{\pm k'}]= [P_{\pm k}, P_{\mp k'}]=0$$
for $k\ne k'$. Thus $P_{+k}$ annihilates all eigenstates of $H_{\rm IM}$ that do not have the $+\epsilon_k$ contribution to their eigenvalue, and likewise $P_{-k}$ annihilates all eigenstates with the $-\epsilon_k$ contribution.
Therefore the eigenstate of $H_{\rm IM}$ with energy $E=\sum_k \alpha_k \epsilon_k$ is
$$
P_{\alpha_1k_1}P_{\alpha_2k_2}\cdots P_{\alpha_Lk_L}|R\rangle
$$
for any choice of $\alpha_a=\pm 1$ and any reference state $|R\rangle$ (say all spins up) not annihilated by any of the $P_{\pm k}$.  Therefore the Hamiltonian must be
\be
H_{\rm IM}=\sum_{k=1}^L \epsilon_k(P_{+k}-P_{-k}) \ .
\label{HPff}
\ee

\newcommand{\Qt}[1]{\widetilde{Q}_{#1}}

A similar rewriting will provide the direct solution of Baxter's clock Hamiltonian. However, there it is not obvious how to work out the analog of the algebra (\ref{Cliffk}) directly in general. Thus I give an alternate derivation of (\ref{HPff}), by a generalization of the brute force ``Bogoliubov transformation''. This involves rewriting $\psi_a$ in terms of the $\Psi_{\pm k}$, plugging it into $H_{\rm IM}$, and then manipulating the resulting expression using (\ref{Cliffk}).  With translation invariance, the rewriting of $\psi_a$ is simply an inverse Fourier transformation. Here the situation is only slightly more complicated, but it is not really: since the $\Psi_{\pm k}$ are built from the eigenvectors of a hermitian matrix, the rewriting simply amounts to a unitary change of basis of the Hilbert space. 

The raising and lowering operators in terms of the $\psi_a$ are
\be \Psi_{\pm k} = \frac{1}{2}\sum_{a=1}^{2L} (\pm i)^a\, \Qt{a-1}(\epsilon_k^2)\,\psi_{a} \ ,
\label{Psiff}
\ee
because the eigenvectors of the matrix $\cal F$ with eigenvalues $\pm \epsilon_k$ are 
\be
\mu_{a}= (\pm 1)^{a}\, \Qt{a-1}(\epsilon_k^2)\ ,
\label{evF}
\ee
where $\Qt{a}$ is a rescaled version of $Q_a$ defined by
$$\Qt{a}(u)=Q_a(u)\prod_{b=a+1}^{2L-1} t_b\  \ .
$$
This can be inverted to write $\psi_a$ in terms of the $\Psi_{\pm k}$ by using orthonormality relations for the $\Qt a$. 
Namely, the orthogonality of the eigenvectors with different eigenvalues requires 
\be
\sum_{j=0}^{L-1} \Qt{2j}(u_k)\Qt{2j}(u_{k'}) = \sum_{j=0}^{L-1} \Qt{2j+1}(u_k)\Qt{2j+1}(u_{k'}) = N_k \,\delta_{k,k'}\ \ .
\label{ortho}
\ee
The normalizations are such that (\ref{Psiff}) indeed satisfies (\ref{Cliffk}) with the same $N_k$.
A real symmetric matrix ${\cal F}$ is diagonalized taking $U {\cal F} U^T$, where $U$ is an orthogonal matrix formed from the normalized eigenvectors. The relations (\ref{ortho}) follow from $U U^T=1$, while $U^T U=1$ implies
\be
\sum_{k=1}^{L} \frac{1}{N_k}\Qt{2j}(u_k)\Qt{2j'}(u_k) = \delta_{j,j'}\ ,\qquad
\sum_{k=1}^{L} \frac{1}{N_k}\Qt{2j+1}(u_k)\Qt{2j'+1}(u_k) = \delta_{j,j'}\ 
\label{ortho1}
\ee
Using this with (\ref{Psiff})
then gives
\be
\psi_{2j} = (-1)^j \sum_{k=1}^L \frac{i}{N_k}\Qt{2j-1}(\epsilon_k^2)\left(\Psi_{+k} + \Psi_{-k}\right)\ ,\quad
\psi_{2j+1} = (-1)^j \sum_{k=1}^L \frac{1}{N_k}\Qt{2j}(\epsilon_k^2)\left(\Psi_{+k} - \Psi_{-k}\right)\ 
\label{psiPsi}
\ee
Plugging this into (\ref{HIF}) and using the orthogonality relations (\ref{ortho}) indeed results in (\ref{HPff}).

\subsection{Higher Hamiltonians for free fermions}
\label{sec:higherff}

As shown in section \ref{sec:bogo}, the Hamiltonian must be written as (\ref{HPff}) because no state can be annihilated by both $\Psi_{+k}$ and $\Psi_{-k}$. This makes it easy to find ``higher'' Hamiltonians whose eigenvalues are given by (\ref{higherspectrum}). These are 
\be
H^{(m)}_{\rm IM}=\sum_{k=1}^L \epsilon_k^m\, (P_{+k}+(-1)^m P_{-k}) 
\label{Hmff}
\ee
where the projectors are defined in (\ref{projop}).
For $m$ even, this is trivially true because $P_{+k}+P_{-k}=1$, as follows from (\ref{Cliffk}). 
Rewritten in terms of the fermions, these are
\be
H^{(m)}_{\rm IM}= \sum_{b=1}^{2L-1}\sum_{\ww=1,3,\dots}^{m}  i^\ww\, \psi_{b+\ww} ({\cal F}^{m})_{b+\ww,b}\, \psi_{b}
\label{HmM}
\ee
for $m$ odd.
The upper limit on sum over $\ww$ arises because $({\cal F}^m)_{b+\ww,b}$ vanishes for $\ww>m$. Hence the higher Hamiltonians are indeed local. 

To prove (\ref{HmM}), first note that given the explicit expression (\ref{Psiff}) for the fermionic shift operators, the projectors  are 
\be
P_{\pm k}=\frac{1}{N_k}\Psi_{\pm k}\Psi_{\mp k} =\frac{1}{2}\pm \frac{1}{2}
\sum_{b=1}^{2L-1}\sum_{\ww=1,3,5,\dots} \frac{i^\ww}{N_k} \Qt{b+\ww-1}(\epsilon_k^2)\Qt{b-1}(\epsilon_k^2)
\,\psi_{b+\ww}\psi_{b}
\label{PIexp}
\ee
by using (\ref{ortho}) for the first piece and setting $\psi_a=0$ for $a>2L$.  The ``width'' $\ww$ of the bilinear is always odd as a consequence of the fermionic anticommutation relation $\{\psi_a,\psi_b\}=2\delta_{ab}$.  Next, an orthogonality relation involving powers of $\epsilon_k$ is needed. For example, using  (\ref{Qrecur}) with (\ref{ortho1}) gives
\be
\sum_{k=1}^L \frac{\epsilon_k}{N_k}\, \Qt{2j-2}(\epsilon_k^2)\Qt{2j'-1}(\epsilon_k^2) = t_{2j'}\delta_{j,j'+1} + t_{2j-1}\delta_{j,j'} = {\cal F}_{2j-1,2j'}
\label{ortho2}
\ee
using the matrix ${\cal F}$ defined in (\ref{calFdef}). Using this with (\ref{HPff}) and (\ref{PIexp}) for $m=1$ gives indeed $H_{\rm IM}$ from (\ref{HIF}).
The orthogonality relation involving higher powers of $\epsilon_k$ is easiest to find by noting that ${\cal F}^2$ has two eigenvectors for each eigenvalue $\epsilon_k^2$ and that ${\cal F}^2$ splits into blocks acting on even and odd indices independently. Thus using (\ref{evF}), these eigenvectors can be taken to have entries $(1+(-1)^a)\Qt{a-1}(\epsilon_k^2)$ 
and $(1-(-1)^a)\Qt{a-1}(\epsilon_k^2)$ in the even and odd blocks respectively. Thus
\be
 \sum_{j'=1}^L({\cal F}^2)_{2j,2j'}\Qt{2j'-1}(\epsilon_k^2)=\epsilon_k^2\, \Qt{2j-1}(\epsilon_k^2)\ ,\qquad
\sum_{j'=1}^L({\cal F}^2)_{2j-1,2j'-1}\Qt{2j'-2}(\epsilon_k^2)=\epsilon_k^2\, \Qt{2j-2}(\epsilon_k^2)\ .
\label{evFsq}
\ee
Since (\ref{ortho1}) and (\ref{ortho2}) hold for all $j$ independent of $j'$, they can be multiplied by the even and odd blocks of ${\cal F}$ respectively. This gives the desired powers of $\epsilon_k$ so that
whenever $(-1)^m = (-1)^{a+a'}$,
\be
\sum_{k=1}^L \frac{\epsilon_k^m}{N_k}\, \Qt{a-1}(\epsilon_k^2)\Qt{a'-1}(\epsilon_k^2) =  ({\cal F}^m)_{a,a'}\ .
\label{orthom}
\ee
Using this with (\ref{PIexp}) and (\ref{Hmff}) results in (\ref{HmM}).

\section{Baxter's clock Hamiltonian and its conserved quantities}
\label{sec:zn}

In this section I describe Baxter's clock Hamiltonian for ${\mathbb Z}_n$-invariant clock chains, and construct both non-local conserved quantities and a sequence of commuting Hamiltonians.
 
\subsection{Baxter's clock Hamiltonian}

A clock chain has an $n$-state quantum system on each site, i.e.\ for $L$ sites the Hilbert space is $(\mathbb{C}^n)^{\otimes L}$. The basic operators $\sigma$ and $\tau$ acting on a single site generalize the Pauli matrices $\sigma^x$ and $\sigma^z$ to an $n$-dimensional space. Instead of anticommutation relations, the operators satisfy
\begin{eqnarray}
\label{znspin1}
\sigma^n = \tau^n =1\ , \qquad \sigma^\dagger &=& \sigma^{n-1}\ , \qquad \tau^\dagger = \tau^{n-1}\ ,\\
\sigma\tau &=& \omega\, \tau\sigma\ ,
\label{znspin2}
\end{eqnarray}
where $\omega\equiv e^{2\pi i/n}$.  
It can be useful to keep in mind an explicit representation of these operators. Diagonalizing one (say $\sigma$) gives
\be \sigma = 
\begin{pmatrix}
1&0&0&\ \cdots\  & 0\\
0&\omega&0&\ \cdots\  & 0\\
0&0&\omega^2&\   & 0\\
\vdots&\vdots&&&\vdots \\
0&0&0&\ \cdots\  & \omega^{n-1}
\end{pmatrix},\quad\quad
 \tau = 
\begin{pmatrix}
0&0&0\ \cdots\  &0& 1\\
1&0&0\ \cdots\  &0 & 0\\
0&1&0\ \cdots   & 0 &0\\
\vdots&&&\vdots&\vdots \\
0&0&0\ \cdots\ & 1& 0
\end{pmatrix}
\label{explicitrep}
\ee
In this representation $\sigma$ generalizes the Pauli matrix $\sigma^z$ to measure the value of a clock variable, while $\tau$ generalizes $\sigma^x$ to shifting the value. The operators $\sigma_j=1\otimes \cdots 1\otimes \sigma \otimes 1\cdots $ and $\tau_j=1\otimes \cdots 1\otimes \tau \otimes 1\cdots $ then act non-trivially at site $j$ of the chain. Each pair $(\sigma_j,\,\tau_j)$ satisfies the algebra (\ref{znspin1},\ref{znspin2}), while operators at different sites commute. 

Baxter's clock Hamiltonian \cite{Baxterclock} is simply
\be
H = \sum_{j=1}^L t^{}_{2j-1}\, \tau_j 
\ +\ \sum_{j=1}^{L-1} t^{}_{2j}\, \sigma_j^\dagger\sigma_{j+1}
 \label{Hclock}
\ee
where the $t_a$ are non-zero real couplings. It is invariant under the $\zn$ symmetry of sending $\sigma_j \to \omega \sigma_j$ for all $j$, i.e.\ shifting each spin. The corresponding symmetry generator, generalizing $(-1)^F$,  is
\be
\omega^{\cal P}\equiv \prod_{j=1}^L \tau_j\ ,
\label{wP}
\ee
which indeed satisfies $(\omega^{\cal P})^n_{} =1$.

This is the simplest possible generalization of the Ising Hamiltonian (\ref{HIsing}): the one-site term generalizes the flip term, whereas the two-site term generalizes the nearest-neighbor interaction. For $n>2$, this Hamiltonian is not hermitian -- I have not accidentally omitted the h.c.! Its eigenvalues hence need not be real. In fact, for any eigenvalue $E$, there must be another eigenvalue $\omega E$, generalizing the invariance of the Ising spectrum under $E\to -E$. This follows from the unitary transformation generalizing charge conjugation in the Ising case to $H\to {\cal C} H {\cal C}^\dagger=\omega H$ here, where
\be
{\cal C} = \left(\prod_{j=1}^L \sigma_j\right)\left( \prod_{k=1}^L \tau_k^{-k}\right)\ .
\label{Cdef}
\ee
Since $H$ is not hermitian, there is no guarantee that it has a complete set of eigenvectors and eigenvalues. Nonetheless, Baxter showed by an indirect method (taking limits of a two-dimensional classical model) that it has indeed $n^L$ eigenvalues given by the remarkably simple formula  (\ref{spectrum}).

With a $\ztwo$ symmetry in the couplings Baxter's clock Hamiltonian is $PT$-symmetric. Spatial parity is implemented by an operator $P$ obeying $P^2=1$ and
$$ P \sigma_j P = \sigma_{L+1-j}\ ,\qquad P\tau_j  P = \tau_{L+1-j}\ .$$
When acting on the Hilbert space  ${\mathbb C^n}^{\otimes L}$, $P$ reverses the order of the $n$-state systems.
Time-reversal invariance is implemented by an operator $T$ obeying $T^2=1$ obeying
$$ T \sigma_j T = \sigma_{j}^\dagger,\qquad T\tau_j  T = \tau_{j}\ .$$
For this to be consistent with the relation $\sigma_j\tau_j =\omega \tau_j \sigma_j$, $T$ must be {antiunitary} (in the basis given by \ref{explicitrep}) as well, so that $\sigma_j\tau_j=T(\sigma_j^\dagger\tau_j)T=T(\overline{\omega}\tau_j\sigma_j^\dagger)T={\omega} \tau_j\sigma_j$ as needed. The Hamiltonian (\ref{Hclock}) is obviously not invariant under these operations individually, but is invariant under $PT$ as long as the couplings are real and obey $t_{2L-a}=t_a$. 
\comment{It is worth noting that by this definition of $T$, the Ising chain is time-reversal invariant, even away from the critical point. However, it is often stated (correctly) that in the Kitaev chain (the Ising chain where the physical degrees of freedom are the fermions \cite{KitMajorana}), time-reversal invariance is broken away from criticality.  This seeming contradiction is resolved by noting that in the Kitaev chain, the action of time-reversal invariance is fixed by assuming that the fermionic operators create actual fermions, where $T^2=-1$.  
}

As described for a more general clock Hamiltonian in \cite{para}, $H$ can be rewritten in terms of parafermions, $\zn$ generalizations of the Majorana fermions discussed above \cite{FK}. Although this mapping is not essential to the subsequent analysis, it turns out to be quite useful intuitively.   At each site of the $n$-state clock chain, there are two basic parafermions 
\be
\psi_{2j-1} =\left(\prod_{k=1}^{j-1} \tau_k\right) \sigma_j \ ,\qquad
\psi_{2j} =\omega^{(n-1)/2}\,\psi_{2j-1}\tau_j\ .
\label{psidef}
\end{equation}
Like $\sigma$ and $\tau$, these do not square to 1 and do not commute, but rather
$$
(\psi_a)^n =1\  , \qquad \psi_a^\dagger = (\psi_a)^{n-1}\ ,\qquad \psi_a\psi_b = \omega\, \psi_b\psi_a\ ,
\hbox{ for } a<b.
$$
The restriction $a<b$ is necessary for these relations to make sense; only for $n=2$ is $\omega=\omega^{-1}$.  
Baxter's clock Hamiltonian (\ref{Hclock}) can be written simply in terms of parafermions:
$$
H=\omega^{(n-1)/2}\sum_{a=1}^{2L-1}t_{a} \psi^{}_{a+1}\psi_a^\dagger\ . 
$$

\subsection{Non-local conserved quantities}
\label{subsec:conserved}

Here I display and discuss the non-local conserved quantities for Baxter's clock Hamiltonian, a result useful for the analysis of the spectrum. 

It is convenient to label each term in the Hamiltonian as $h_a$ for $a=1\dots 2L-1$, so that 
\be H= \sum_{a=1}^{2L-1} h_a \ ,\ee
i.e.\ $h_{2j-1}=t_{2j-1}\tau_j$ 
and $h_{2j}=t_{2j}\sigma^\dagger_j\sigma_{j+1}$. 
The $h_a$ obey the algebra
\begin{equation}
h_a h_{a+1} = \omega\, h_{a+1} h_a\ ,\qquad (h_a)^n=(t_a)^n\equiv \gamma_a\ ,\qquad [h_a,\, h_b] = 0 \quad\hbox{for}\quad |a-b|>1\ .
\label{hcomm}
\ee
The $h_a$ therefore generate a $n^{2L-1}$-dimensional algebra similar to that of the parafermions, but here only ``adjacent'' ones pick up a factor of $\omega$ on reordering. A very special and extremely important property of this Hamiltonian is that $h_a^n=\gamma_a$ is proportional to the identity.

Non-local conserved quantities are straightforward to find explicitly. Note that as a consequence of (\ref{hcomm}), $[h_a, h_{a-1}h_{a+1}]=0$. 
This suggests defining
\begin{equation}
J^{(2)} = \sum_{c=b+2}^{2L-1}\sum_{b=1}^{2L-3} h_c\, h_b
\label{Jtwo}
\ee
as the sum of all bilinears in the $h_a$ such that the two are always at least two sites apart. The proof that $[H,\, J^{(2})]=0$ is easy to do by direct computation. However, it is more illuminating and general to prove that $H$ commutes with an entire hierarchy of non-local conserved charges obeying the same exclusion rule, where the $h_a$ must be at least two apart. Namely, define
\be
J^{(m)}_{2L} = \sum_{b_m=b_{m-1}+2}^{2L-1} \cdots \sum_{b_2=b_1+2}^{2L-(2m-3)}\,\sum_{b_1=1}^{2L-(2m-1)} h_{b_m}\cdots h_{b_2}h_{b_1}
\label{Jdef}
\ee
for $m=1\dots L$. The Hamiltonian for $L$ sites is by definition $J^{(1)}_{2L}$. The ``highest'' charge is
\be
J^{(L)}_{2L}= h_{2L-1}\cdots h_3h_1 =  \omega^{\cal P} \prod_{j=1}^L  t_{2j-1}
\label{JL}
\ee
where $\omega^{\cal P}$ is the $\zn$ conserved charge defined in (\ref{wP}).

A not-really coincidence to note is that the non-local charges satisfy the same exclusion rule as the polynomials $Q_a$ in (\ref{Qdef}) whose roots give the $\epsilon_k$. They therefore satisfy similar recursion relations, namely 
\be
J^{(m)}_{a+1} = J^{(m)}_{a}\ +\ h_a\,J^{(m-1)}_{a-1}\ ,
\label{Jrecur}
\ee
where $J^{(0)}_a=1$ and $J_a^{(m)}$ is defined by replacing $2L$ with $a$ in $(\ref{Jdef})$. Since (\ref{Jrecur}) relates the $J^{(m)}_a$ for different $a$, recursion can be used to prove that $J^{(m)}$ commutes with $H=J^{(1)}_{2L}$ at all system sizes. Making the standard recursive assumption that $[J^{(1)}_{a},\, J^{(m)}_a]=0$ for all $m$ and for all $a$ below some fixed value yields
\begin{eqnarray*}
[J^{(1)}_{a+1},\, J^{(m)}_{a+1}] &=&[J^{(1)}_a+h_a,\ J^{(m)}_a + h_a J^{(m-1)}_{a-1}] \\
&=&[h_a,J^{(m)}_a]\ +\ [J^{(1)}_a,\, h_a J^{(m-1)}_{a-1}]\\
&=&[h_a,\, h_{a-1} J_{a-2}^{(m-1)}]\ +\ [h_{a-1},\, h_aJ^{(m-1)}_{a-1}]\\
&=&[h_a,\, h_{a-1}] J_{a-2}^{(m-1)}\ +\ [h_{a-1},\, h_a(J^{(m-1)}_{a-2}+h_{a-2} J^{(m-2)}_{a-3})]\\
&=&[h_{a-1},\, h_{a}h_{a-2}  J^{(m-2)}_{a-3}]\\
&=&0
\end{eqnarray*}
by repeatedly using the recursion relation and the fact that $[h_a,J^{(m)}_{b}]=0$ for all $b\le a-1$. This therefore proves by recursion that for all $L$
\be 
[H, J^{(m)}_{2L}]= 0\ .
\ee
The same sort of recursive proof shows that all the non-local conserved charges commute:
\be 
[J^{(m)}_a, J^{(m')}_{a}]= 0\ .
\label{Jcomm}
\ee

With this Hamiltonian, the Hilbert space and the operators on it can be graded in a useful fashion. Namely, consider the operator
\be
G=\prod_{j=1}^L\left(\omega^{-j(n-1)/2}\sigma_j\tau_j^{-j}\right)\ .
\label{Gdef}
\ee
This operator obeys $G^n=1$, and satisfies
\be
Gh_a\ =\ \omega\, h_a G
\ee
for all $a$. Thus if the Hilbert space is organized by the eigenvalues $\omega^g$ of $G$, the operator $H$ shifts $g\to g+1$. Since the conserved charges discussed here are built out of the $h_a$, each of these charges can be labelled by the amount they shift $g$.  The non-local conserved quantity $J^{(m)}$ is built of $m$ powers of the $h_a$, so acting with it on the Hilbert space shifts $g\to g+m$. This superscript is a useful piece of notation, because when multiplying operators it adds mod$\,n$.

\subsection{Higher Hamiltonians}
\label{sec:higher1}

Integrable models typically (always?) have local conserved quantities, where ``local" means that the range of interactions does not increase with the size of the system. Each local conserved quantity thus can be thought of as a Hamiltonian with longer-range interactions. These ``higher'' Hamiltonians commute with $H$ and each other, and so can be diagonalized simultaneously. A remarkable feature of the Baxter clock Hamiltonian is that the higher Hamiltonians have an elegant explicit form. In this subsection and in appendix \ref{app:higher} I derive these higher Hamiltonians. These results apply for arbitrary $\omega$; in the next section \ref{sec:Hwn} I describe their special properties when $\omega^n=1$.

In a classical integrable model, the higher Hamiltonians are found from commuting transfer matrices \cite{Baxbook}. In an integrable two-dimensional classical model, the transfer matrix $T$ can be written in terms of a particular combination of couplings $u$ so that $[T(u),\,T(u')]=0$. The Hamiltonian and the higher Hamiltonians of the corresponding integrable quantum chain are then defined by taking the logarithmic derivative and expanding the result in a power series:
\be
-u\frac{d}{du} \ln T(u) = \sum_{m=1}^{\infty} H^{(m)} u^m = H u + H^{(2)}u^2 +H^{(3)}u^3 + \dots 
\label{higherH}
\ee
Because the transfer matrices at different $u$ commute, all the higher $H$ must commute:
\be
[H^{(m)},\,H^{(m')}] =0\ .
\label{Hcomm}
\ee
where the Hamiltonian itself is the first one: $H=H^{(1)}$.

The non-local conserved quantities in this quantum integrable model can be combined into an object very much like a commuting transfer matrix, namely
\be 
T(u) = 1 - J^{(1)}u + J^{(2)}u^2 + \dots + (-1)^L J^{(L)}u^L = \sum_{m=1}^L  (-u)^m J^{(m)}\ .
\label{TJ}
\ee
This sum truncates because there are no $J^{(m)}$ for $m>L$. Because all of the $J^{(m)}$ commute among themselves, this results in  $[T(u),\,T(u')]=0$. Defining $H^{(m)}$ via the logarithmic derivative (\ref{higherH}) of $T(u)$ from (\ref{TJ}) yields a set of commuting higher Hamiltonians. 
Taking the derivative and expanding out (\ref{higherH}) gives each $H^{(m)}$ in terms of lower ones:
\be
H_a^{(m)} = (-1)^{m+1} m\, J_a^{(m)} + \sum_{q=1}^{m-1} (-1)^{q+1} H_a^{(m-q)} J_a^{(q)}
\label{HJ}
\ee
When the subscript on $H^{(m)}$ is omitted, this always means $H_{2L}^{(m)}$.

Explicit expressions for the first few higher Hamiltonians can be worked out by brute force by using the commutation relation (\ref{hcomm}), giving
\begin{eqnarray} 
\nonumber
H^{(1)}_a&=& H\ =\ \sum_{b=1}^{a-1} h_b\\
\label{H2}
H^{(2)}_a&=& \sum_{b=1}^{a-1}[ h_b^2\ +\ (1+\omega) h_{b+1} h_{b}]\\
\nonumber
H^{(3)}_a&=& \sum_{b=1}^{a-1}[ h_b^3\ +\ (1+\omega+\omega^2)(h_{b+2}h_{b+1}h_{b} + h_{b+1}^2h_b + h_{b+1}h_b^2)]
\end{eqnarray}
where for ease of notation all $h_b$ with $b>a-1$ are defined to be zero. The explicit calculation of $\hhh{2}$ is given in this section, while that for $\hhh{3}$ is in appendix \ref{app:explicit}. One might be tempted to guess a general expression from these, but the answer is not so simple: for example, the coefficient of $h_{b+1}^2h_b^2$ in  $H^{(4)}$ turns out to be $(1+\omega)(1+\omega+\omega^2)$. 

It is plausible but not immediately apparent that the $H^{(m)}$ defined in this fashion are local. However, it is simple to derive that not only are the higher Hamiltonians local, but {\em connected} as well. This means that any time $h_b$ and $h_{b+b'}$ appear in a term, all those in between (i.e.\ $h_{b_i}$ with $b<b_i<b+b'$) appear as well with non-vanishing power. This automatically implies locality because there are exactly $m$ of the $h_{b_i}$ in each term. 

Connectedness is equivalent to having $D^{(0,m)}=0$ for $m>0$, where
$D^{(s,m-s)}$ is the piece of $H_a^{(m)} -H_{a-1}^{(m)}$ proportional to $h_{a-1}^s$, i.e.\ 
\be
\sum_{s=0}^m D^{(s,m-s)}= H_a^{(m)} - H_{a-1}^{(m)}\ ,\qquad
D^{(s,m-s)} h_a = \omega^s\, h_a D^{(s,m-s)} \ .
\label{Ddef}
\ee
This is proved by plugging (\ref{Ddef}) into the definition (\ref{HJ}) of $\hh{m}{a}$, and using the recursion relation (\ref{Jrecur}) to make all dependence on $h_{a-1}$ explicit. 
\comment{giving
\begin{eqnarray*}
H_{a-1}^{(m)}+ \sum_{s=0}^{m}D^{(s,m-s)}
 &=& \sum_{q=1}^{m-1}(-1)^{q+1}\Big(H^{(m-q)}_{a-1}
  + \sum_{s=0}^{m-q} D^{(s,m-q-s)}\Big)(J^{(q)}_{a-1}+h_{a-1}J_{a-2}^{(q-1)})\\
    &&\qquad +\ (-1)^{m+1} m (J_{a-1}^{(m)}+h_{a-1} J^{(m-1)}_{a-2})\ .
\end{eqnarray*}
}
Collecting the terms with no $h_{a-1}$ present gives
$$H_{a-1}^{(m)}+ D^{(0,m)} = (-1)^{m+1}m\,J_{a-1}^{(m)}+\sum_{q=1}^{m-1}(-1)^{q+1}(H^{(m-q)}_{a-1}+D^{(0,m-q)}) J^{(q)}_{a-1}\ .
$$
Using the definition (\ref{HJ}) for $H^{(m)}_{a-1}$ leaves this to be
$$D^{(0,m)} = \sum_{q=1}^{m-1}(-1)^{q+1}D^{(0,m-q)} J^{(q)}_{a-1}
$$
for all $m\ge 1$. This means that $D^{(0,1)}=0$, as is easy to check from the definition. Thus
all vanish:
\be
D^{(0,m)} = 0 
\ .
\label{D0m}
\ee
A useful fact shown in section \ref{sec:Hwn} is that an even stronger definition of connectedness holds, namely that $H^{(m)}$ remains connected even after simplifying using the fact that $h_b^n=\gamma_n$ is a number.

There is a closed-form expression for all the $H^{(m)}$. 
It is easy to check using (\ref{HJ}) that
\be
D^{(m,0)}=h_{a-1}^m\ .
\label{Dmm}
\ee
The rest of them can be determined recursively.
Trial and error and the construction of the shift operators in section \ref{sec:shift} suggests the recursion relation
\be
H^{(m)}_{a+1}=H_{a}^{(m)}+ h_a^m +  \sum_{r=1}^{m-1}\sum_{s=1}^{m-r} \frac{\beta_m}{\beta_{m-r}}
 h_a^r A_{rs} D^{(s,m-r-s)}\ ,
\label{ansatz}
\ee
for some coefficients $\beta_m$ and the $A_{rs}$. 

To check the recursion relation for $m=2$, note that (\ref{HJ}) gives
$$
H^{(2)}_{a+1}=(H_a^{(1)}+h_a)(J_{a}^{(1)}+h_a)-2(J^{(2)}_a+h_aJ^{(1)}_{a-1})\ .
$$
Using (\ref{HJ}) for $H^{(2)}_a$ and $J^{(1)}_a=H^{(1)}_a$ gives
$$H^{(2)}_{a+1}=\hh{2}{a}+h_a^2+
[H_a, h_a] + 2h_a(H_{a}-H_{a-1})= H_{a}^{(2)}+h_a^2 + (1+\omega)h_ah_{a-1}
$$
This indeed yields the explicit expression (\ref{H2}). 
The ansatz reads
$$H^{(2)}_{a+1}=H_{a}^{(2)}+h_a^2 + \beta_2 h_a A_{11} D^{(1,0)}\ .$$
Since the $\beta_m$ appear in ratios, they can be rescaled to set $\beta_1=1$. 
Because $D^{(1,0)}=h_{a-1}$, this gives
\be
\beta_2 A_{11}= 1+\omega\ .
\label{b2A}
\ee

By considering some special cases, all the coefficients $\beta_m$ and $A_{rs}$ can be fixed. This calculation is done in appendix \ref{app:coeff}. 
The $\beta_m$ are given by
\be
\beta_m =\frac{1-\omega^m}{1-\omega}
\label{betam}
\ee
for $m\ge 1$. Each $A_{rs}$ is a {\em Gaussian binomial} (aka q-binomial):
\be
A_{rs} = \frac{(1-\omega^{r+1})(1-\omega^{r+2})\dots (1-\omega^{r+s-1})}{(1-\omega)(1-\omega^2)\dots(1-\omega^{s-1})}
=  \frac{(1-\omega^{s})(1-\omega^{s+1})\dots (1-\omega^{r+s-1})}{(1-\omega)(1-\omega^2)\dots(1-\omega^{r})}\ .
\label{Ars}
\ee
The first of the two forms requires $s>1$, while the second requires $r>0$, so $A_{0s}=1$ follows from the first while $A_{r1}=1$ follows from the second. The proof that the ansatz with these coefficients gives the correct $H^{(m)}$ defined by (\ref{HJ}) is given in appendix \ref{app:proof}. 

This results in a closed-form expression for $\hhh{m}$. Each term is of the form 
$$\bb{m}{r_{1}} A_{r_Wr_{W-1}}\dots A_{r_3r_2} A_{r_2r_1}h_{c+W-1}^{r_W}
\dots h_{c+1}^{r_2}\dots h_{c}^{r_1}\ .$$
Because $\hhh{m}$ by construction contains $m$ powers of the $h_a$, $\sum_{j=1}^W r_j =m$, while 
because of the connectedness, each $r_j>0$. All such terms appear so
\be
\hh{m}{a}= \sum_{c=1}^{a-1} \sum^{(m)}\,\bb{m}{r_1}
\prod_{j=1}^{W} A_{r_{j+1}r_{j}}h_{c+j-1}^{r_j}\ 
\label{Hexp}
\ee
where $\sum^{(m)}$ is the sum over all $r_j\ge 1$ and $W$ such that $\sum_{j=1}^W r_j = m$. Setting $r_{W+1}=0$, is harmless to include $A_{r_{W+1}r_W}=A_{0r_W}=1$ in the product.  This and all products of operators in this paper are ordered so that the $j=1$ term in the product is the {\em rightmost} one. 

It is straightforward to check that $[H,\, H^{(m)}]=0$ directly using this expression; in fact, this is shown in appendix \ref{app:proof} as a part of the proof of (\ref{Hexp}).  The benefit of defining the higher Hamiltonians using (\ref{HJ}), rather than simply taking (\ref{Hexp}) as the definition of the $\hhh{m}$, is twofold. Since the explicit expression is obviously unwieldy, the definition in terms of the conserved charges gives a broader picture of the structure of the theory. The second reason is more practical: with the definition (\ref{HJ}),  the higher Hamiltonians automatically commute. Proving $[\hhh{m},\hhh{m'}]=0$ using the explicit expression is a combinatorial nightmare.

The explicit expression (\ref{Hexp}) does not look at all like the expression (\ref{HmM}) for the higher Hamiltonians in the free fermion case. Even ignoring surface dissimilarities, there is a fundamental difference: the number of terms in (\ref{Hexp}) grows exponentially with system size, where the number of fermion blinears in (\ref{HmM}) grows only quadratically.


\section{The Hamiltonians for $\omega^n=1$}
\label{sec:Hwn}

The construction of the higher Hamiltonians in section \ref{sec:higher1} applies to any $\omega$. The purpose of this section is to derived the interesting consequences of setting $\omega^n=1$, so that Baxter's Hamiltonian is for a clock variable. For example, when $m$ is a multiple of $n$,  $\beta_m=0$, and it seems as if higher Hamiltonians  $\hhh{sn}$ vanish. As I will explain in section \ref{sec:higher2}, this is not quite so but rather they are proportional to the identity. Even more interesting is that the constant of proportionality reveals a deep connection to the free-fermion case. 
In fact, there are deep connections to the free-fermion case for all $m$. In section \ref{sec:Hrewritten} I show that $\hhh{m}$ can be written in a form resembling (\ref{HmM}). This rewriting will prove essential in the construction of the shift and projection operators in the next sections. The beautiful thing about the resulting formula (\ref{HMx}) is that virtually all of the remaining analysis will be a variation on that for free fermions.


The key technical observation making the results of this section possible is that the coefficients $A_{rs}$ from (\ref{Ars}) in the explicit expression (\ref{Hexp}) for $\hhh{m}$ have a very simple behavior when the labels are shifted by $n$. Namely,  
consider the range $0\le r < n$, $1\le r'\le n$ where there are manifestly no singularities in $A_{rr'}$ in the limit $\omega^n\to 1$. Then for $l$ and $l'$ any non-negative integers,
\be
\frac{A_{r+ln,r'+l'n}}{A_{rr'}}=
\frac{(l+l')!}{l!\,l'!}\equiv \binom{l+l'}{l}\ .
\label{Arln}
\ee
An important result is that this ratio is not only independent of $r$ and $r'$, but of $n$ as well.

\subsection{The Hamiltonians $\hhh{sn}$}
\label{sec:higher2}

An important first thing to check is that there are no singularities in the explicit expression (\ref{Hexp}) for $\hhh{m}$ in the limit $\omega^n\to 1$. The $A_{rs}$ are all finite because all zeroes of the denominator are cancelled by zeroes in the numerator, a consequence of (\ref{Arln}). However, because $\beta_{sn}=0$, the $\beta_{r_{1}}$ in the denominator would result in a singularity if not cancelled by a zero of one of the other coefficients. The explicit expression (\ref{Ars}) shows that
$$\frac{A_{r_2 r_1}}{\beta_{r_1}}=\frac{A_{r_1r_2}}{\beta_{r_2}}\ .$$
Since all $A_{rs}$ are finite, a singularity in (\ref{Hexp}) therefore can only occur at $\omega^{{\rm gcf}(r_{2},r_{1})}=1$,
where gcf$(r,r')$ means the greatest common factor of $r$ and $r'$. Moving the $1/\beta_{r_{j}}$ down the line in this fashion means that if there is a singularity, it can only occur at $\omega^{g}=1$, where $g={\rm gcf}(r_1,r_2,\dots ,r_W)$. However, the overall factor $\beta_m$ multiplying the coefficient will cancel any such singularity because $m=\sum_{j=1}^W r_j$ is necessarily a multiple of $g$. Thus the limit of $\hhh{m}$  in (\ref{Hexp}) as $\omega^n\to 1$ is not singular, i.e.\ the explicit expression can still be used if interpreted as a limit. 

This presence of the $\beta_{r_1}$ in the denominator does mean that $H^{(sn)}$ is not automatically zero despite a $\beta_{sn}=0$ in front of (\ref{Hexp}). This is obvious because of the $h_b^{sn}$ term, which always occurs with coefficient one. This is not the only non-vanishing term in general: if $g$ is a multiple of $n$, the zero in $\beta_{sn}$ is cancelled. The only way for this to happen if {\em all} the $r_j$ are multiples of $n$. Since $h_b^{rn}=\gamma_n^r$, $\hhh{sn}$ is therefore proportional to the identity. 

There is an elegant explicit expression for $\hhh{sn}$. 
Because $A_{0n}=1$, (\ref{Arln}) implies that for $r\ge 0$ and $r'\ge 1$, 
\be A_{ln,l'n}=\frac{(l+l'-1)!}{l!\,(l'-1)!}= \lim_{\omega\to 1} A_{l,l'}\ ,
\label{Arnsn}
\ee
i.e.\ when both coefficients are multiples of $n$ the Gaussian binomial turns into an ordinary one. 
Similarly,
$$\lim_{\omega^n\to 1} \bb{sn}{ln}=\frac{s}{l}=\lim_{\omega\to 1} \bb{s}{l} \ .$$
Since all non-vanishing terms in (\ref{Hexp}) for $\hhh{sn}$ have $r_j$ proportional to $n$, letting $r_j=nl_j$ means that the constraint $\sum_{j=1}^W r_j=m$ turns into $\sum_{j=1}^W l_j =s$, and so
\begin{eqnarray*}
\hh{sn}{a}&=& \sum_{c=1}^{a-1}\sum^{(sn)} \bb{sn}{nl_1} 
\prod_{j=1}^{W} A_{nl_{j+1},nl_{j}}h_{c+j-1}^{nl_j}\\
&=& \lim_{\omega \to 1} \, \sum_{c=1}^{a-1} 
\sum^{(s)}
\bb{s}{l_1} \prod_{j=1}^{W} A_{l_{j+1},l_{j}}\gamma_{c+j-1}^{l_j}\\
&=& \lim_{\omega \to 1}\, \widetilde{H}^{(s)}_a\ ,
\end{eqnarray*}
where $\widetilde{H}^{(s)}_a$ is $\hh{s}{a}$ with the $h_{c+j-1}$ replaced with $\gamma_{c+j-1}$. Since the $\hhh{m}$ are generated by the logarithmic derivative (\ref{higherH}) of $T(u)$, the $\widetilde{H}^{(s)}_a$ can be generated in a similar fashion. Since all $h_{b}$ commute with each other when $\omega\to 1$,  the $\hh{sn}a$ are given by (\ref{higherH}) with the same replacements of $h$ with $\gamma$ in $T(u)$. Strikingly, $T(u)$ with this replacement has already appeared above, in the solution of the free-fermion problem in section \ref{sec:Ising}! Precisely, in terms of the function $Q$ in (\ref{Qdef}),
\be
\sum_{s=1}^\infty \hh{sn}{a} u^s= -u\frac{\partial}{\partial u} \ln \left(u^{a/2}\, Q_a\left(u^{-1}\right)\right)\ .
\label{HQ}
\ee
This is why it was convenient to define $Q_a$ in terms of the $\gamma_b$, which for general $n$ are related to the couplings via $\gamma_b=t_b^n$.

This is the first indication how the solution of Baxter's clock Hamiltonian is very closely related to the free-fermion solution of the Ising chain. An even more striking relation comes from a fundamental result in the theory of symmetric polynomials, the Newton-Girard formula (yes, that Newton) \cite{Mathworld}. Define the symmetric polynomials ${\cal S}_l$ and $e_l$ via
$$ {\cal S}_l (u_1,\,\dots u_L)= \sum_{k=1}^L u_k^l\ ,\qquad\quad
e_l (u_1,\,\dots u_L)= \sum_{k_1<k_2<\dots k_l\le L} u_{k_1} u_{k_2}\dots u_{k_l} \ .
$$
Then it is simple to prove recursively that
\be
(-1)^s s\, e_s + \sum_{l=0}^{s-1} (-1)^{l} {\cal S}_{s-l} \,e_{l}^{} = 0\ .
\label{NG}
\ee
Note that this is the same relation (\ref{HJ}) relating the non-local conserved charges and the higher Hamiltonians, so it can be generated by a logarithmic derivative as well. Using it with (\ref{HQ}) gives $\hh{sn}{a}$ in terms of the roots of $Q_{2L}$. Namely, letting these roots be $u_k$ for $k=1\dots L$ gives 
\begin{eqnarray*}
Q_{2L}(u) &=& \prod_{k=1}^{L} (u-u_k)
=\sum_{l=0}^L u^{L-l} (-1)^l e_{l}(u_1,\,\dots u_L) \ .
\end{eqnarray*}
The Newton-Girard formula (\ref{NG}) combined with (\ref{HQ}) then gives
\be
\hh{sn}{a}= \sum_{k=1}^L u_k^s\ .
\label{Hsn}
\ee
For $n=2$, these roots obey $u_k=\epsilon_k^2$, where the $\epsilon_k$ are precisely the energy levels of the free-fermion problem. In section \ref{sec:shift} I rederive Baxter's result that for general $n$ the energy levels are related to these same roots with the identification $u_k=\epsilon_k^n$. 
Thus 
\be
\hh{sn}{a}= \sum_{k=1}^L \epsilon_k^{sn}\ 
\label{Hsne}
\ee
for all $n$.

\newcommand{\xb}[2]{\xi^{(#1)}_{#2}}
\newcommand{\Xq}[2]{\Xi^{(#1)}_{#2}}
\newcommand{\xmw}{\xi^{(m)}_{\ww,b}}
\newcommand{\fmw}{f_{s,\ww,b}}

\subsection{A useful basis}
\label{sec:reduced}

Fermion systems are free when their Hamiltonian is a sum of fermion bilinears. Quite obviously the terms in the higher Hamiltonians (\ref{Hexp}) for general $n$ are not as simple as a fermion bilinear. However, the general Hamiltonians can be rewritten so that not only are they considerably simpler, but also emphasize how much of the structure of these models is independent of $n$. 

The first step in doing so is to find a useful basis in which to write the operators. The higher Hamiltonians satisfy a number of useful properties that make this possible. Namely, after ``reducing'' (\ref{Hexp}) by using $h_c^n=t_c^n \equiv\gamma_n$ to reduce the all the exponents in (\ref{Hexp}) to less than $n$, two useful properties of the resulting expression are:
\begin{itemize}
\item An {\em exclusion rule}: adjacent exponents must sum to at most $n$: i.e.\ $r_j+r_{j+1}\le n$.
\item {\em Connectedness}: each term remains connected.
\end{itemize}

The ``exclusion'' rule is a consequence of the fact that
the coefficient $A_{rr'}$ from (\ref{Ars}) obeys
\be
A_{r+ln,r'+l'n}=0\quad \hbox{ when }\  r +r'  > n\ ,
\label{Arsexc}
\ee
with $0\le r< n$ and $1\le r'\le n$.
Thus only terms where 
\be
r_j\,\hbox{mod}\,n\ +\ r_{j+1}\,\hbox{mod}\,n \le n
\label{exc}
\ee
appear in $\hhh{m}$. For $n=2$, this means only fermion bilinears appear in the Hamiltonians. For $n=3$, this means for example that $h_{b+1}^2h_b^2$ never appears.

The stronger form of connectedness is a consequence of (\ref{Arsexc}), the fact that
$$ A_{r+ln,l'n}=0\quad\hbox{ for }r\ne 0\ . $$
This shows that $h_{b}^{n}=\gamma_{b}$ to any non-zero power cannot appear in the ``middle'' of a term. Namely, any term in $H^{(m)}$ obeys
\be
\begin{cases}
r_j\,\hbox{mod}\,n\ne 0 \qquad&\hbox{ in the ``middle'' region }\  b-c< j \le b-c+\ww\\
r_j=s_j n \hbox{ with }s_j>0 \qquad&\hbox{ in the ``outer'' region }\  j\le b-c\hbox{ or } j> b-c+\ww\ 
\end{cases}
\label{rbb}
\ee
for some $0\le \ww\le W$ and $c\le b< c+W$. This property arises because even though $A_{r,l'n}=0$ for $r\,$mod$\,n\ne 0$, the exponent $r_W$ on the left end can be a multiple of $n$ because $A_{nl,s}\ne 0$. If $r_W$ is a multiple of $n$, then $r_{W-1}$ can be as well because $A_{rn,rs}$ is also non-vanishing, and so on. The same holds starting from the other end by using $\beta_rA_{rs}=\beta_sA_{sr}$, i.e.\ if $r_{1}$ is a multiple of $n$, then $r_{2}$ can be as well. However, (\ref{Arsexc}) means that if $r_{j}\,$mod$\,n\ne 0$ and $r_{j'}\,$mod$\,n\ne 0$, then
all $r_{j''}$ in between the two (i.e. with  $j<j''<j'$) also obey $r_{j''}\ne 0\,$mod$\,n$. 
Thus once a term is reduced, it still remains connected even ignoring any $\gamma_a$.

These properties and (\ref{Arln}) mean that $\hhh{m}$ can be written as a linear combination of {\em reduced, fixed-width, operators}. These operators are defined to have no $\gamma_a$, and are
\be
\xb{m}{\ww,b}= {\sum^{(m,\ww)}}\,\prod_{j=1}^\ww\bb{m}{r_1}A_{r_{j+1}r_{j}}h_{b+j-1}^{r_j}\ ,
\label{xmw}
\ee
where $\sum^{(m,\ww)}$ is the sum over all $1\le r_j <n$ such that $\sum_{j=1}^\ww r_j = m$ and $r_{\ww+1}=0$. This differs from $\sum^{(m)}$ in that there is no sum over widths and that each $r_j$ is restricted to be less than $n$. The restriction means no $\gamma_a$ appear in any of the $\xb{m}{\ww,b}$.
It is convenient to define $\xb{m}{0,b}=\delta_{m0}$ as well.

The $\xmw$ have some general characteristics. As described in section \ref{sec:higher2}, when $m$ a multiple of $n$, $\beta_m=0$, so the Hamiltonian is proportional to the identity. Here this manifests itself by making $\xb{sn}{\ww,b}=0$ unless $\ww=s=0$. Other $\xmw$ vanish as well. As a consequence of connectedness, $\xb{m}{\ww,b}=0$  for $\ww>m$. Less obvious but important is that, as a consequence of the restriction and the exclusion rule (\ref{exc}), $\xb{m}{\ww,b}=0$ for $\ww<2[m/n]+1$ as well.  Conversely, for a fixed value of $\ww$, the maximum value of $m$ allowed is $n[(\ww+1)/2]-1$. Since the maximum value of $\ww$ is $2L-1$, this means the maximum value of $m$ in any $\xb{m}{\ww,b}$ is $nL-1$.

The Hamiltonians for any $n$ can be written as a sum over the $\xb{\ww}{b}$. This comes from using (\ref{Arln}) to relate any coefficients to those with indices $r_j<n$ and $h_a^n=\gamma_a$ to do the same to the exponents. The fact that these ratios depend only $[r_j/n]$ means that 
\be
H^{(m)} =  \sum_{b=1}^{2L-1}\sum_{\ww=1}^{2L-1}  \sum_{s=0}^{[m/n]}
 \fmw\, \xb{m-sn}{\ww,b}\ .
 \label{Hf}
\ee
While the coefficients $\fmw$ in this sum in principle could depend on $m$ and $n$, in fact when written in terms of the $\gamma_a$ they do not. This characteristic is essential to all that follows. I show this by deriving an ugly but explicit expression for them, and use this in the next subsection \ref{sec:Hrewritten} to find a much nicer one.

Combining the expression (\ref{Hexp}) for $\hhh{m}$ with the definition (\ref{xmw}) of the reduced, fixed-width operators and defining $s_j=[r_j/n]$ gives
$$\fmw=
\sum^{(s)}{}'\,
\left(\prod_{j=b-c+\ww+1}^{W} A_{r_{j+1}r_{j}}\gamma_{c+j-1}^{s_j}\right)
\left(\prod_{j=b-c+1}^{b-c+\ww} \frac{A_{r_{j+1}r_j}}{A_{r_{j+1}{\rm mod}\, n,\,r_{j}{\rm mod}\, n}}
\gamma_{c+j-1}^{s_j}\right)
\left(\prod_{j=1}^{b-c}A_{r_j r_{j+1}}\gamma_{c+j-1}^{s_j}\right)
$$
where $\sum^{(s)'}$ means the sum over $r_j$ and $W$ such that $\sum_{j=1}^W [r_j/n]=s$ while imposing the additional restrictions (\ref{rbb}) on  the $r_j$. In terms of the $s_j$, these restrictions are $s_j\ge 1$ in the outer region. The reversed order of the subscripts in the last product is not a typo, but is a consequence of using $\beta_rA_{rs}=\beta_sA_{sr}$ to turn the $\beta_{r_1}$ into a $\beta_{r_b}$ to conform with the definition of $\xmw$. Using (\ref{Arln}) gives
\be \fmw = \sum^{(s)}{}'\
\prod_{j=b-c+\ww+1}^{W-1} \binom{s_j+s_{j+1}-1}{s_{j+1}}
\prod^{b-c+\ww}_{j=b-c} \binom{s_j+s_{j+1}}{s_j}
\prod^{b-c-1}_{j=1} \binom{s_j+s_{j+1}-1}{s_j}
\prod_{j=1}^{W}\gamma_{c+j-1}^{s_j}\ .
\label{fbinom}
\ee
Ugly though this expression is, it does not depend in any way on $m$ or $n$, but rather only on $s$.  In the next section I find a much simpler formula for it.

It is worth noting that even though $\xmw$ is much more complicated that a fermion bilinear, it has parafermions at the ends. Namely, using the definition of parafermions given in (\ref{psidef}) gives
$h_b \propto \psi_a^\dagger$, and so
\be
\xmw  \propto\ \psi_{b+{\rm w}} \psi^\dagger_b \ 
\ee
multiplied in general by additional powers of $h_{b_j}$ with $b\le b_j\le b+\ww$.
For $n=3$, the exclusion rule requires that all terms must be proportional to
$$ \psi_{b+\ww}\, h_{b_l}\, \dots h_{b_1}\, \psi^\dagger_b $$
where ${b+\ww}\ge b_j \ge h_b$ and $b_{j+1}-b_{j} >1$, i.e.\ the $h_{b_j}$ here cannot be adjacent. 
\comment{
For the free-fermion case $n=2$, the exclusion rule mean the only terms are products of this type, so all terms in the higher Hamiltonian are simply fermion bilinears. Likewise, for higher $n$, terms have $\psi_{b+\ww}$ on one end and $\psi_b^\dagger$ on the other, with $h_{b_j}$ in between.
}

\subsection{The Hamiltonians rewritten}
\label{sec:Hrewritten}

The free-fermion higher Hamiltonians were rewritten in terms of fermion bilinears in (\ref{HmM}).
Here I prove an analogous formulae for all $n$, namely
\be
H^{(m)} =  \sum_{b=1}^{2L-1} \sum_{\ww=0}^{2L-1}\sum_{s=0}^{[m/n]}
 ({\cal M}^{2s+\ww })_{b+\ww,b}\, \xb{m-sn}{\ww,b}\ ,
 \label{HMx}
\ee
where the matrix ${\cal M}$ is a rescaling of the free-fermion matrix ${\cal F}$:
\be
{\cal M}=\begin{pmatrix}
0&\gamma_1&0&\cdots\\
1&0&\gamma_2&  \\
0&1&0&\\
\vdots&&&&\ 0&\gamma_{2L-1}\\
&&&&\  1&0
\end{pmatrix}\ ,
\label{calMdef}
\ee
i.e.\ the matrix elements are $({\cal M})_{ab}=\delta_{a,b-1}+\gamma_a\delta_{a,b+1}$.
Given how complicated the $\hhh{m}$ are in their original definition, this formula is remarkably simple. Note in particular that ${\cal M}$ does not depend on $n$ or $m$ at all, but only on the couplings $\gamma_a$; the only dependence on $m$ and $n$ is via the reduced, fixed-width operators $\xmw$.

To give a simple example, consider $\hhh{n+1}$, where the only terms involving a $\gamma_a$ have $\ww=1$. Using $A_{1n}/\beta_n= A_{n1}/\beta_1=A_{n1}=A_{11}=1$ gives
\begin{eqnarray*}
\hhh{n+1}&=& \sum_{b=1}^{2L-1}\left( h_b^{n+1}+ h_{b+1}^n h_b + h_{b+1}h_b^n +
 \sum_{\ww=3}^{n+1} \xb{n+1}{\ww,b}\right)\\
 &=&\sum_{b=1}^{2L-1}\left( (\gamma_{b-1}+\gamma_b+\gamma_{b+1})\xb{1}{1,b} +
 \sum_{\ww=3}^{n+1} \xb{n+1}{\ww,b}\right)
\end{eqnarray*}
The coefficient $\gamma_{b-1}+\gamma_b+\gamma_{b+1}$ is indeed $({\cal M}^3)_{b+1,b}$\,, while ${\cal M}_{b+1,b}=1$. 

The proof of (\ref{HMx}) for all $n$ comes almost entirely by examining the free-fermion case, because the explicit expression (\ref{fbinom}) for $\fmw$ is independent of $n$. In the $n=2$ case there are already two very different-looking expressions for the higher Hamiltonians. The general expression (\ref{Hexp}), and its expansion (\ref{Hf}) in terms of the $\xmw$ still applies for $n=2$, while (\ref{HmM}) gives the higher Hamiltonians in terms of fermion bilinears. Thus for $m$ odd a funny equality holds
\be
\hhh{m}\Big|_{n=2} =i \sum_{b=1}^{2L-1}\sum_{\ww=1,3\dots}^{m} (-1)^\ww\psi_{b+\ww} ({\cal F}^{m})_{b+\ww,b}\, \psi_{b} =  \sum_{b=1}^{2L-1}\sum_{\ww=1,3,\dots} f_{(m-\ww)/2,\ww,b}\, \xb{\ww}{\ww,b}
\label{HmHm}
\ee
There is a single term $s$ obeying $m=2s+\ww$, because there is only non-vanishing $\xmw$ for $n=2$ for each $m$, namely $\xb{\ww}{\ww,b}$. The $\xb{\ww}{\ww,b}$ are indeed fermion bilinears:
$$\xb{\ww}{\ww,b}= h_{b+\ww-1}\dots h_{b+1}h_{b}= \psi_{b+\ww} \psi_b\,\prod_{l=1}^{\ww} (it_{b+l})\ ,$$
since $A_{11}=A_{01}=1$ and $A_{10}=0$. Absorbing these factors of $t$ into a rescaling of ${\cal F}$ into ${\cal M}$ from (\ref{calMdef}) gives (for $\ww$ odd)
\be
({\cal M}^{2s+\ww})_{b+\ww,b}=  f_{s,\ww,b}\ .
\label{Mf}
\ee

Neither side of (\ref{Mf}) depends in any way on $n$ -- this is nothing more nor nothing less than a very interesting combinatorial identity. The ugly expression (\ref{fbinom}) turns out to simplify remarkably!
It is easy to check explicitly that the identity holds when  $\ww$ is even as well.  Plugging (\ref{Mf}) into (\ref{Hf}) then yields the simple expression (\ref{HMx}) for the higher Hamiltonians.

\section{Shift operators}

\label{sec:shift}

In this section I explicitly  construct the ``shift" operators, the analog for Baxter's clock Hamiltonian of the raising and lowering operators for the Ising chain. Instead of raising or lowering the energy like $\Psi_{\pm k}$ do in the free-fermion case, acting with a shift operator changes the energy by a complex number. In particular, for a chain of $L$ sites, there are $nL$ shift operators $\Psi_{\omega^l,k}$ with $l=0\dots n-1$ and  $k=1\dots L$. These obey 
\be
[H,\,\Psi_{\omega^p,k}] =(1-\omega) \omega^p \epsilon_k\, \Psi_{\omega^p,k}\ ,
\label{Hpsik}
\ee
where $\epsilon_k$ is real. 
Thus $\Psi_{\omega^l,k}$ shifts one of the ``free parafermions'' in figure \ref{fig:paralevels} clockwise.

In addition, I show that commuting the same shift operators with the higher Hamiltonians gives 
\be
[\hhh{m},\, \Psi_{\omega^p,k}] = \epsilon_k^m \omega^{pm} (1-\omega^m)\,\Psi_{\omega^p,k}
\label{Hmshift}
\ee
Thus $\Psi_{\omega^p,k}$ behaves as a shift operator for all the Hamiltonians, except when $m$ is a multiple of $n$, where $\hhh{m}$ is proportional to the identity so that the commutator vanishes.
This gives a strong indication that all the Hamiltonians have spectrum (\ref{higherspectrum}).

\subsection{Commuting with $H$ and the higher Hamiltonians}
\label{sec:comH}

The key to the construction of the rasing/lowering operators in Ising is that commuting the Hamiltonian with something linear in the fermions gives something linear in the fermions. It would have been nice if the same were true of the parafermions defined in (\ref{psidef}), but as described in detail in \cite{para}, this is not the case. 
Calculating a few commutators explicitly illustrates the problem needing to be solved. The simplest parafermion is $\psi_1\equiv\sigma_1$, and its commutator with $H$ is also simple:
$$[H,\psi_1]= (1-\omega) h_1\psi_1\ \propto 
\psi_2
$$
using the fact that $h_a\psi_1=\omega^{\delta_{a1}}\psi_1 h_a\ .$
However, it starts to get ugly quickly:
$$[H,[H,\,\psi_1]]=(1-\omega)^2h_2h_1\psi_1 + (1-\omega)^2 h_1^2\psi_1\ , $$
$$[H,[H,[H,\,\psi_1]]] = (1-\omega)^3h_3h_2h_1\psi_1 + (1-\omega)^3 h_2^2h_1\psi_1\ 
+ (1-\omega)^2(1-\omega^2)h_2h_1^2\psi_1 + (1-\omega^3)h_1^3\psi_1\ .$$

The key to making progress is to observe that the terms here look very similar to the higher Hamiltonians appearing in (\ref{H2}). This suggests trying to write the commutators in terms of commutators with $\hhh{m}$.  Using the explicit forms from in (\ref{H2}) indeed gives
\begin{eqnarray*}&&[H,[H,\psi_1]]=\frac{(1-\omega)^2}{1-\omega^2} [\hhh{2},\psi_1]\ ,\\ 
&&[H,[H,[H,\psi_1]]]=\frac{(1-\omega)^3}{1-\omega^3} [\hhh{3},\psi_1]\ .
\end{eqnarray*}
This correspondence is true in general. In appendix  \ref{app:shift} I prove that when the commutator is done $m$ times,
\be 
[H,[H,\cdots, [H,\, \psi_1]\cdots ]] =\lim_{\omega^n\to 1}\frac{(1-\omega)^m}{1-\omega^m} [\hhh{m},\psi_1]\ .
\label{Hmpsi}
\ee
The reason this needs to be defined as a limit is that the denominator vanishes for $m$ a multiple of $n$. This limit makes sense as a consequence of the properties of $\hhh{sn}$ discussed in detail in section \ref{sec:higher2}. All terms are either proportional to $\beta_{sn}$ or $h_1^{nl_1}$, so in the former, $\beta_{sn}$ cancels the denominator here, while in the latter the commutator with $\psi_1$ gives a factor of $(1-\omega^{nl_1})$, again canceling the zero in the denominator. Thus even though $\hhh{sn}$ itself is proportional to the identity, commuting $sn$ times with $H$ gives a non-trivial operator.

It is convenient to define an operator ${\cal H}$ which acts on operators:
\be {\cal H}X\equiv \frac{1}{1-\omega} [H_{2L},X]\ .
\ee
i.e.\ operating with ${\cal H}$ corresponds to commuting with $H_{2L}$ and dividing by $1-\omega$. (In the math literature, ${\cal H}$ would be called $\hbox{ad}_{H_{2L}/(1-\omega)}\,$.) Thus (\ref{Hmpsi}) is
 \be
 {\cal H}^m\psi_1 = \lim_{\omega\to e^{2\pi i /n}}\frac{[\hh{m}{2L},\psi_1]}{1-\omega^m} \ .
 \label{Hs}
 \ee
The matrix ${\cal F}$ in (\ref{calFdef}) is simply ${\cal H}$ for $n=2$ acting in the basis given by $\psi_1,\dots, \psi_{2L}$. Because $H$ commutes with the $\zn$ charge $\omega^{\cal P}$, all $\HH^m\psi_1$ have $\zn$ charge $-1$, i.e. $\omega^{\cal P}\HH^m\psi_1\omega^{-\cal P}
=\omega^{-1}\HH^m\psi_1$.

\subsection{The shift operators and a $\zn$ analog of the Majorana fermions}
\label{sec:vn}

The shift operators obeying (\ref{Hpsik}) are simply the eigenvectors of ${\cal H}$.
As shown in the previous subsection, repeatedly acting with $\HH$ (i.e.\ commuting with $H$) on $\psi_1$ produces a tractable set of operators with a closed-form expression. The hope is then that this sequence of commutators effectively truncates, so that acting with $\HH$ produces only some linear combination of commutators already seen. For example, in the fermion case $\HH^{2L}\psi_1$ gives some linear combination of the operators ${\cal H}^m\psi_1$ with $m<2L$. This means the fermion raising/lowering operators are constructed simply by diagonalizing a $2L\times 2L$ matrix. 

\comment{A hint of where it happens comes from the exclusion rule (\ref{exc}), which requires that $\hh{Ln}{2L}$ must have at least one $h_{b}^{r}$ with $r>n$: there is no way that the product of $2L-1$ different $h_{b}$ can have exponents $r_b < n$ that obey both $\sum_b r_b=Ln$ and the exclusion rule. Moreover, the explicit formulae shows that  if one ignores the $\gamma_b$, all terms in $\hh{Ln}{2L}$ appear in some lower $\hh{sn}{2L}$.
}


Such a truncation does indeed occur for all $n$, i.e.\ $\HH^m\psi_1$ for $m\ge nL$ is given by a linear combination of those with $m<nL$.
The proof is straightforward given the explicit expression for the Hamiltonians, and is given in appendix \ref{app:shiftproof}. Here I define operators $\vv{m}$ so that  ${\cal H}$ can be written in a matrix form analogous to (\ref{psiprime},\ref{calFdef}). 
Defining $\vz\equiv\psi_1$, $\vv{m}$ in the free-fermion case is proportional to $\psi_{m+1}$, but for general $n$ it will necessarily be more complicated.

The construction relies on the fact that when commuting with $H$ repeatedly to generate the $\vv{m}$, one finds that lower ones appear multiplied by various $\gamma_a$. These then can be subtracted off to define subsequent $\vv{m}$. The first $\gamma_a$ appears in $\HH^n\vz$:
$${\cal H}^n \vz\  =\ h_{1}^{n}\psi_1 + \dots\ =\ \gamma_1\vz + \dots\ .$$
So this suggests defining 
\be
\vv{l} = {\cal H}^l \vz\  \hbox{ for } l<n
\label{vl0}
\ee
but
$$
\vv{n} = \HH^n \vz - \gamma_1\vz\ .
$$
In the free fermion-case $n=2$ this indeed results in $\vv{1}\propto \psi_2$ and $\vv{2}\propto \psi_3$. Similarly, it follows from the explicit 
expressions (e.g.\ (\ref{Hpsi})) that 
$$\HH \vv{n} = \gamma_2 \psi_2 + \dots$$
so it is sensible to define
$$
\vv{n+1}= \HH \vv{n} - \gamma_2 \HH \vz\ .$$
Continuing in this fashion, all $\vv{a}$ are defined by (\ref{vl0}) and
\begin{eqnarray}
\nonumber
\vv{sn} &=& \HH \vv{sn-1}\ -\ \gamma_{2m-1} \vv{sn-n}\ ,\\
\label{vHv}
\vv{sn+1} &=& \HH \vv{sn}\  -\ \gamma_{2m}\, \vv{sn-n+1}\ ,\\
\nonumber
\vv{sn+l+1} &=& \HH \vv{sn+l}
\end{eqnarray}
with $m\ge 1$ and $l=1,\dots n-2$. When $n=2$, this indeed reduces to $\vv{m}\propto \psi_{m+1}$. Note that the $\gamma_a$ vanish for $a>2L-1$, so the last non-trivial subtraction occurs in $\vv{Ln}$.
This recursive definition of the $\vv{a}$ is very similar to that of the recursion relation (\ref{Qrecur}) for the functions $Q_a(u)$ used to solve the free-fermion problem, and in fact
\be \vv{sn} = Q_{2s}(\HH^n)\vz\ ,\qquad 
\qquad \vv{sn+l} = \HH^{-n/2}Q_{2s+1}(\HH^n)\vv{l}\  \quad\hbox{for }1\le l<n\ .
\label{vQ}
\ee
This is because the recursion relations (\ref{vHv}) imply e.g.\
\be
\vv{sn+n}=(\HH^n -\gamma_{2s+1}-\gamma_{2s})\vv{sn}+\gamma_{2s+1}\gamma_{2s-1}\vv{sn-n}\ ,
\label{vrecur}
\ee
while from (\ref{Qrecur}) it follows that
\be
Q_{a+2}(u) = (u -\gamma_{a+1}-\gamma_{a})Q_a(u)+\gamma_{a+1}\gamma_{a-1}Q_{a-2}(u)\ .
\label{Qrecur3}
\ee

Although  $\HH^m\vz$ is non-zero for all $m$, $\vv{m}$ is not. As proved in appendix \ref{app:shiftproof}
in (\ref{QHzero}), 
\be
\vv{nL} = Q_{2L}(\HH^n)\vz = 0\ .
\label{vLn}
\ee
It is also possible to prove this by explicitly working out the action of $\HH$ on the $\vv{m}$, but this is quite tedious. One finds that $\vv{sn}$ has a minimum width in the sense that if $h_{2s}$ is set to zero, then $\vv{sn}=0$. Because $h_{2L}=0$, it follows that $\vv{nL}=0$. 

With this truncation and the definition (\ref{vHv}),  $\HH$ indeed acts linearly on a finite set of the $nL$ operators $\vv{m}$, just like for free fermions. The energy levels $\epsilon_k$ therefore follow simply from the eigenvalues of the resulting $nL \times nL$ matrix. Defining the linear combination
$$\Psi= \sum_{a=0}^{nL-1} \mu_a \vv{a}{}$$
means
\be
\HH\Psi = \sum_{a=0}^{nL-1} \mu'_a \vv{a}\qquad\hbox{where }\ \mu'_{a}=\sum_{b=0}^{nL-1} ({\cal M}_n)_{ab}\, \mu_b\ .
\label{HPsi}
\ee
The eigenvalues of ${\cal H}$ are then those of the matrix ${\cal M}_n$.
The matrix elements of ${\cal M}_n$ follow from the definition of the $\vv{a}$:
\be
({\cal M}_n)_{b+1,b}=1,\qquad ({\cal M}_n)_{sn,sn+n-1}=\gamma_{2m-1},\qquad 
({\cal M}_n)_{sn+1,sn+n}=\gamma_{2m}\ ,
\label{Mndef}
\ee
with all others vanishing. For $n=2$, this is precisely the matrix ${\cal M}$ defined in (\ref{calMdef}). 
For $n=3$ it is
$${\cal M}_3=
\begin{pmatrix}
0&0&\gamma_1&0&0&0&0&\cdots\\
1&0&0&\gamma_2&0&0&0&\cdots\\
0& 1& 0&0&0&0&0&\\
0&0&1&0&0& \gamma_3& 0&\\
0&0&0&1&0&0&\gamma_4\\
\\
\vdots&\vdots
\\
&&&&&&&\ 0&0&\gamma_{2L-1}\\
&&&&&&&\ 1&0&0\\
&&&&&&&\ 0&1&0
\end{pmatrix}\ .
$$

The eigenvalues can be easily found by noting that because of the grading, the matrix $({\cal M}_n)^n$ breaks into $n$ blocks of size $L\times L$. Namely, by the definition (\ref{vHv}), $\HH\vv{sn+q}$ always gives some combination of operators $\vv{s'n+q+1}$. Thus ${\cal H}^n$ mixes only $\vv{s'n+q}$ with the same value of $q$. Then the resulting blocks ${\cal B}_q$ are independent of $n$, and are 
\be
({\cal B}_0)_{aa'}  = ({\cal M}^2)_{2a-1,2a'-1},\qquad ({\cal B}_l)_{aa'}  = ({\cal M}^2)_{2a,2a'}\ .
\label{BM}
\ee
with $l=1,\dots n-1$. 
Thus no more work is necessary to find the eigenvalues -- the values $\epsilon_k^n$ for general $n$ are indeed the same as the $\epsilon_k^2$  in the free-fermion case! This confirms Baxter's result
\be
Q_{2L}(\epsilon_k^n)=0\ .
\label{Qroots}
\ee
This result is rederived in a slightly different fashion in appendix \ref{app:shiftproof}. In the special case where all couplings are equal so that $\gamma_a$ is independent of $a$, these eigenvalues are $\epsilon^n_k = 2\cos(\pi k/(2L+1))$.

An explicit expression for the shift operators in terms of the $\vv{m}$ follows directly from (\ref{HPsi}) by computing the eigenvectors of the matrix ${\cal M}_n$. The eigenvectors of $\HH^n$ with eigenvalue $u_k=\epsilon_k^n$ are
\be
\Phi^{(0)}_k =\frac{1}{N_k}\sum_{m=0}^{L-1}   \QQ{2m}(u_k)\, \vv{sn}\ ,\qquad
\Phi^{(l)}_k= \frac{1}{N_k}\epsilon_k^{n/2}\sum_{m=0}^{L-1} \QQ{2m+1}(u_k)\, \vv{sn+l}\ ,
\label{chidef}
\ee
for $l=1\dots n-1$. The polynomial $\QQ{a}$ is a rescaled version of $Q_a$:
\be
\QQ{a}(u_k)=Q_{a}(u_k)\prod_{b=a+1}^{2L-1} \gamma_b\ .
\label{QQdef}
\ee
The action of $\HH$ on the $\Phi^{(q)}$ is simple to work out using (\ref{vHv}); it is obvious that $\HH\Phi^{(l)}_k=\Phi^{(l+1)}_k$ for $l=1,\dots n-2$, while
\begin{eqnarray*}
\HH\Phi^{(0)}_k &=& \frac{1}{N_k} \sum_{m=0}^{L-1} \QQ{2m}(u_k)\,
\left(\vv{sn+1}+\gamma_{2m-1}\, \vv{sn-n+1}\right)\\
&=&\frac{1}{N_k} \sum_{m=0}^{L-1} \left(\prod_{b=2m+2}^{2L-1}\gamma_b\right)\left(\gamma_{2m+1}Q_{2m}(u_k) +Q_{2m+2}(u_k)\right)\vv{sn}
\\
&=& \Phi^{(1)}_k
\end{eqnarray*}
using $Q_{2L}(u_k)=0$ and the recursion relation (\ref{Qrecur}). Similarly, $
\HH\Phi^{(n-1)}_k =
\epsilon_k^n\,\Phi^{(0)}_k$.

The shift operators are linear combinations of the $\Phi^{(q)}_k$.
Since $u_k$ is the square of the eigenvalue of a hermitian matrix ${\cal F}$, it is real and positive, and so $\epsilon_k=(u_k)^{1/n}$ can be taken real and positive as well. Then
\be
\Psi_{\omega^p,k} = \sum_{q=0}^{n-1} (\omega^p\epsilon_k)^{-q}
\Phi^{(q)}_k(\epsilon_k^n)\ .
\label{Psired}
\ee
indeed satisfy
\be
\HH\Psi_{\omega^p,k}=\omega^p \epsilon_k \Psi_{\omega^p,k}\ .
\label{HPsiepsi}
\ee
The $\Phi^{(q)}_k$ can be written in terms of the $\Psi_{\omega^p,k}$ by the obvious inverse discrete Fourier transform.

\comment{
This indeed yields 
$$
\Psi_{\omega^{p},k}= \sum_{q=0}^{n-1} \left(\omega^{p} \epsilon_k\right)^{-q} \Phi^{(q)}_k\ 
$$
in agreement with (\ref{Psired}).
}

Using the orthonormality relations (\ref{orthombar}), it is possible to
 ``invert'' the expressions for the shift operators and express the $\vv{a}$ in terms of them:
\be
\vv{sn} = \sum_{m=0}^L  Q_{2m}(u_k) \Phi^{(0)}_k\ ,\qquad 
\vv{sn+q}= \epsilon_k^{-n/2}\sum_{m=0}^L Q_{2m+1}(u_k)  \Phi^{(l)}_k\ .
\label{vchi}
\ee
for $l=1\dots n-1$.
This is the natural generalization of the relations (\ref{psiPsi}) in the free-fermion case. Thus here and in their commutator with $H$, the $\vv{a}$ behave quite analogously to the fermions. Moreover,  by computer checks it appears that
$$(\vv{m})^n\propto 1\ ,$$
just like the Majorana fermions square to the identity. However, this seems difficult to prove directly, because the commutation relations between different terms in $\vv{m}$ are not nice at all. In section \ref{sec:algebra} I describe further some of the algebraic structure.


The expression (\ref{HPsiepsi}) yields  $nL$ shift operators obeying
(\ref{Hpsik}) as advertised. Acting with $\Psi_{\omega^{p},\epsilon_k}$  on any eigenstate of Baxter's clock Hamiltonian either annihilates it or or shifts it in energy by $(1-\omega)w^{p}\epsilon_k$. 
The fact that the $\epsilon_k^n$ for any $n$ are given by the eigenvalues of the same matrix is precisely Baxter's result \cite{Baxterclock}. 
This is a strong implication that the entire spectrum is given by (\ref{spectrum}), but not quite a proof. I will provide the full demonstration that this is indeed so in section \ref{sec:spectrum}.

\subsection{The higher Hamiltonians and the shift operators}
\label{sec:shiftham}

The higher Hamiltonians provided an essential ingredient in deriving the shift operators. One might expect then that their commutator with the shift operators is very nice, and indeed this is so:
\begin{eqnarray*}
 [\hhh{m},\, \Phi^{(0)}_k]&= & \frac{1}{N_k}\sum_{j=0}^{L-1}  \QQ{2j}(\epsilon_k^n)\,[\hhh{m},\, \vv{jn}]\\
&=& \frac{1}{N_k}\sum_{j=0}^{L-1}  \QQ{2j}(\epsilon_k^n)\,[\hhh{m},\, Q_{2j}(\HH^n)\,\vz]\\
&=& (1-\omega^m)\frac{1}{N_k}\sum_{j=0}^{L-1}  \QQ{2j}(\epsilon_k^n) Q_{2j}(\HH^n)\HH^m \vz\\
&=&  (1-\omega^m)\frac{1}{N_k}\sum_{j=0}^{L-1}  \QQ{2j}(\epsilon_k^n)
 \frac{1}{n}\sum_{k'=1}^{2L}\sum_{p=0}^{n-1}  Q_{2j}(\epsilon_{k'}^n)
(\epsilon_{k'}\omega^p)^m \,\Psi_{\omega^p,{k'}}\\
&=&(1-\omega^m) \epsilon_{k}^m\frac{1}{n}\sum_{p=0}^{n-1}\omega^{pm}\Psi_{\omega^p,{k}}
\end{eqnarray*}
using
(\ref{chidef}), (\ref{vQ}), the fact that $[H,H^{(m)}]=0$, (\ref{Hs}), (\ref{Hshift}), and (\ref{ortho}). Similarly, 
$$[\hhh{m},\Phi^{(l)}_k] = (1-\omega^m) \frac{1}{n}\sum_{p=0}^{n-1}(\epsilon_{k}\omega^{p})^{m+l}\,\Psi_{\omega^p,{k}}
$$
for $l=1,\dots n-1$. Note that these commutators vanish for $m$ a multiple of $n$, a consequence of $\hhh{sn}$ being a multiple of the identity. Using these with (\ref{Psired}) gives (\ref{Hmshift}), 
$$
[\hhh{m},\, \Psi_{\omega^p,k}] = \epsilon_k^m \omega^{-pm} (1-\omega^m)\,\Psi_{\omega^p,k}\ ,
$$
as advertised. Thus $\Psi_{\omega^p,k}$ behaves as a shift operator for all the Hamiltonians except when $m$ is a multiple of $n$, where $\hhh{m}$ is proportional to the identity and the commutator vanishes.

This gives a strong indication that all the Hamiltonians have a simple and beautiful spectrum
(\ref{higherspectrum}). Even though the shift operators are not useful for $m$ a multiple of $n$, this formula for the spectrum still applies; this was already shown in (\ref{Hsne}). These results are also a strong hint that not only the commutators but the Hamiltonians themselves can be rewritten in a similar fashion as (\ref{Hshift}). I do this in the next section \ref{sec:spectrum}.

\comment{
which rewritten in terms of $\QQ{a}$ are
\be
\sum_{k=1}^{L} \frac{1}{N_k}Q_{a}(u_k)\QQ{a'}(u_k) = \delta_{a,a'}\ 
\label{orthobar}
\ee

The $\chi_{k,q}$ for all $q=0,\dots, n-1$ can be written in terms of the shift operators as
\be 
\chi_{k,q} = \frac{\epsilon_k^q}{n}\sum_{p=0}^{n-1} \omega^{pq} \Psi_{\omega^p,k}\ .
\ee
}

\section{The spectrum and the algebra}
\label{sec:spectrum}

Since as shown in (\ref{Hpsik}) and (\ref{Hmshift}) there are $nL$ shift operators, it is hard to imagine that the spectrum of the Hamiltonians could be anything other than (\ref{spectrum}) and (\ref{higherspectrum}). 

A simple check on this statement comes from the fact (\ref{JL}) that the non-local conserved quantity $J^{(L)}$ is proportional to the $\zn$ conserved charge $\omega^{\cal P}$, and so can easily be evaluated on any state. Using the Newton-Girard formula (\ref{NG}) with (\ref{spectrum}) and (\ref{higherspectrum}) gives
$$J^{(L)} = \prod_{k=1}^L (\omega^{p_k} \epsilon_k) = \omega^{\sum_k p_k} ({\rm det}{\cal F}^2)^{1/4} = 
 \omega^{\sum_k p_k} \prod_{k=1}^{L} t_{2k-1}\ ,$$
where ${\cal F}$ is our friend from (\ref{calFdef}) whose eigenvalues are $\epsilon_k^2$. This indeed agrees with (\ref{JL}) as long as $\zn$ conserved charge for this energy eigenstate is $\omega^{\cal P} = \omega^{\sum_k p_k}.$ 

Since computing the algebra of the shift operators by brute force does not seem possible, the preceding results do not rule out the possibility that some of the $n^L$ eigenvalues do not appear in the spectrum and some may appear multiple times. This could occur for example if the product $\Psi_{\omega^p,k} \Psi_{\omega^{p'},k'}$ is not proportional to $\Psi_{\omega^{p'},k} \Psi_{\omega^{p},k}$
when $k\ne k'$. Then this would amount to two different operators resulting in the same energy shift, and so the end state could be doubly degenerate. Similarly, if this product vanishes for $k\ne k'$, then this energy shift and the final state may not occur at all.

In this section I rule out these strange possibilities and show that each eigenvalue of the $n^L$ eigenvalues in (\ref{spectrum}) and (\ref{higherspectrum}) does occur exactly once. I do this by rewriting all the Hamiltonians as a sum over commuting operators $\xmw$. The idea is similar to that used to find the projection operators, here using  (\ref{HMx}) to rewrite the Hamiltonian in terms of a simple expression involving the $\epsilon_k$ and reduced, fixed-width operators. The eigenvalues of the $\xmw$ are then worked out by using their commutators with the shift operators and their behavior under charge conjugation.

\newcommand{\Pk}{P^{}_{\omega^p,k}}
\newcommand{\Xk}{\Xi^{(q)}_k}

\subsection{The spectrum}
\label{sec:fullspec}

Defining the reduced, fixed-width operators allowed the Hamiltonians to be written in a form (\ref{HMx}) greatly resembling the Ising expression (\ref{HmM}) in terms of the fermion blinears. This allows the Hamiltonians for {\em all} $m$ to be rewritten in terms of a {\em finite} set of operators $\Xi^{(q)}_k$, with $q=0\dots n-1$ and $k=1\dots L$. The expression is
\be \hhh{m}= \sum_{k=1}^L u_k^{[m/n]}\,\Xi^{(q)}_k\ ,
\label{HX}
\ee
where $q=m\,$mod$\,n$. 
Another nice feature of this expression is that the only dependence on $m$ is via the exponent of $u_k$ and the index $q$.

The proof is simple. The orthonormality relation (\ref{orthombar}) allows the matrix elements of ${\cal M}$  to be rewritten in terms of the polynomials $Q_a$ and their rescaled versions $\QQ{a}$. 
\comment{
$$ \hhh{m}=\sum_{b=1}^{2L-1} \sum_{\ww=0}^{2L-1}\sum_{s=0}^{[m/n]}\sum_{k=1}^L
 u_k^{[m/n]-s+\ww/2}\frac{1}{N_k}\, \QQ{b+\ww-1}(u_k)Q_{b-1}(u_k)\, \xb{ns+q}{\ww,b}\ ,
$$
where $s\to [m/n]-s$ from (\ref{HMx}) and $q=m\,$mod$\,n$. where 
}
Plugging this into (\ref{HMx}) gives (\ref{HX}) with
\be
\Xi^{(q)}_k=\sum_{b=1}^{2L-1} \sum_{\ww=0}^{2L-1}\sum_{s=0}^{s_{\rm max}}
u_k^{-s+\ww/2}\frac{1}{N_k}\, \QQ{b+\ww-1}(u_k)Q_{b-1}(u_k)\, \xb{ns+q}{\ww,b}\ ,
\label{Xdef}
\ee
where $s_{\max}$ is the maximum value of $s$ such that $\xb{sn+q}{\ww}$ is non-vanishing. Its value
follows from the exclusion rule (\ref{exc}), and is explained following (\ref{xmw}). 
What is important is that this does not grow with $m$ except via $q$. 
A check on (\ref{Xdef}) is that 
$$\Xi^{(0)}=1\ ,$$
as follows from the orthonormality relation (\ref{orthobar}) and the fact that $\xb{m}{0,b}= \delta_{m0}$. Since $u_k=\epsilon_k^n$, this now verifies the higher Hamiltonians $\hhh{sn}$ indeed obey the simple form (\ref{Hsne}) for all $n$. Another easily-done check is for $m=1$, where $s_{max}=0$ and $\xb{1}{\ww,b}=\delta_{\ww,1}h_b$. 

Since all Hamiltonians are written in terms of the $(n-1)L$ different $\Xk$,  (\ref{HX}) can be inverted to write the $\Xk$ in terms of the $\hhh{m}$ with $m<nL$. Letting $j=[m/n]$ so that $m=j[m/n]+q$ means (\ref{HX}) is for $j=0\dots L-1$
$$ \hhh{jn+q}= \sum_{k=1}^L {\cal X}_{jk}\, \Xk\ ,$$
where ${\cal X}_{jk}=u_k^{j}$. Written as a matrix, this is
$${\cal X}=
\begin{pmatrix}
1&1&1&\dots&1\\
u_1&u_2&u_3&\dots&u_L\\
u_1^2&u_2^2&u_3^2&\dots& u_L^2\\
\vdots\\
u_1^{L-1}&u_2^{L-1}&u_3^{L-1}&\dots& u_{L}^{L-1}
\end{pmatrix}\ ,
$$
with the Vandermonde determinant: det${\cal X}=\prod_{k<k'}(u_{k'}-u_k)$; it follows from the free-fermion case that this does not vanish for real non-vanishing $t_j$. Thus ${\cal X}$ can be inverted to give
\be
\Xk = \sum_{j=0}^{L-1}({\cal X}^{-1})_{kj}\hhh{jn+q}\ .
\label{XH}
\ee

The consequences of this are enormous. Since the $\hhh{m}$ all commute amongst themselves (why I did all this work in section \ref{sec:higher1} and the appendix), the $\Xk$ do too:
$$
[\Xk,\,\Xq{q'}{k'}] = 0
$$
for all $k,k',q,q'$. 
This means each of the higher Hamiltonians decomposes into the sum of $L$ commuting pieces, and each can be diagonalized individually. Moreover, the pieces have a nice commutation relation with the shift operators:
\begin{eqnarray}
\nonumber
\left[\Xq{q}{k'},\,\Psi_{\omega^{p},k}\right]&=& \sum_{j=0}^{L-1}({\cal X}^{-1})_{k'j}\left[\hhh{jn+q},\,\Psi_{\omega^{p},k}\right]
=\sum_{j=0}^{L-1}({\cal X}^{-1})_{kj}u_k^j(\omega^{p}\epsilon_k)^q(1-\omega^q)\Psi_{\omega^{p},k}\\
&=&\delta_{kk'} 
(\omega^{p}\epsilon_k)^q(1-\omega^q)\Psi_{\omega^{p},k}
\label{XPsi}
\end{eqnarray}
by (\ref{Hmshift}) and the definition of ${\cal X}$. 

All that is left to do is to find the eigenvalues of the $\Xk$.  The fundamental theorem of algebra requires that any square matrix, even those not hermitian, has at least one eigenvector and eigenvalue. Moreover, the $\zn$ generalization of the charge conjugation operator ${\cal C}$ defined in (\ref{Cdef}) has the property ${\cal C}h_b{\cal C}^\dagger = \omega h_b$. Thus
$${\cal C}\,\Xk\,{\cal C}^\dagger = \omega^q \Xk$$
and so for every eigenvalue $\lambda_k$ of $\Xk$, $\omega^q\lambda_k$ is also an eigenvalue. Combining this with (\ref{XPsi}) means that the eigenvalues of $\Xk$ are $(\epsilon_k\omega^{p})^q$. This holds for all $k$, and the fact that det${\cal M}\ne0$ when the couplings $t_b$ are non-zero means that none of the $\epsilon_k$ vanish. Since all $\Xk$ commute among themselves, (\ref{HX}) gives for the spectrum of all the Hamiltonians:
\be 
E^{(m)}=\sum_{k=1}^L (\omega^{p_k} \epsilon_k)^{m} \ ,
\label{highspec}
\ee
i.e.\ (\ref{higherspectrum}). This is the central result of this paper, and reproduces Baxter's result for the original Hamiltonian with $m=1$.

\subsection{The projection operators}
\label{sec:projectors}

The Hamiltonian can be written as the sum over projection operators $\Pk$ as in (\ref{HP}) by defining
\be 
\Pk=\frac{1}{n}\sum_{q=0}^{n-1}(\epsilon_k\omega^{p})^{-q}\,\Xk\ ,\qquad \Xk=\sum_{p=0}^{n-1}(\epsilon_k\omega^{p})^q\,\Pk\ .
\label{Pdef}
\ee
Since all  $\Xk$ commute, the $\Pk$ do as well:
\be
[\Pk,\,P^{}_{\omega^{p'},k'}] =  0
\label{Pcomm}
\ee
for all $k,k',p,p'$. 
Plugging the second part of (\ref{Pdef}) into (\ref{HX}) yields
\be
\hhh{m}=\sum_{k=1}^{L} \sum_{p=0}^{n-1}  (\epsilon_k \omega^p)^m P_{\omega^p,k}\ .
\label{HP}
\ee
For $m$ a multiple of $n$, this already follows from (\ref{Hsne}) because $\sum_{p=0}^{n-1}  P_{\omega^p,k}=\Xi^{(0)}_k =1$. 

Since the spectrum is given by (\ref{highspec}), it is natural to expect that 
$P_{\omega^p,k}$ projects onto states where the $k$th contribution to the energy is $\omega^p \epsilon_k$. The simplest way of seeing this is to note that this is the only consistent way for  (\ref{HP}) to apply for all positive integer $m$ for a finite number of $\Pk$.
This confirms that the shift operators behave as expected.  Given (\ref{highspec}) and the commutator (\ref{XPsi}), the shift operator $\Psi_{\omega^p,k}$ takes the state with eigenvalue $(\omega^{p+1}\epsilon_k)^q$ of $\Xk$ and shifts it to the state with eigenvalue $(\omega^{p}\epsilon_k)^q$. It must annihilate all other eigenstates of $\Xk$, but because of (\ref{Pcomm}) does not affect the eigenvalues of $\Xq{q}{k'}$ with $k'\ne k$. 
More explicitly, using (\ref{Pdef}) with (\ref{XPsi}) 
gives the commutators
$$
\left[P_{\omega^{p'},{k'}}\,,\, \Psi_{\omega^{p},k}\right]=\delta_{kk'}\left(\delta_{p'p}-\delta_{p',p+1}\right)
\Psi_{\omega^{p},k}
$$
Combining this with the above observation gives
\be
P_{\omega^{p'},{k}}\, \Psi_{\omega^{p},k}=\delta_{pp'}\Psi_{\omega^{p},k} \ ,\qquad \Psi_{\omega^{p},k}P_{\omega^{p'},{k}}=\delta_{p',p+1} \Psi_{\omega^{p},k}\ ,
\label{PssP}
\ee
so acting with $\Psi_{\omega^{p},k}$ on some eigenstate of $P_{\omega^{p},{k}}$ shifts its eigenvalue  by $+1$ or annihilates it. 
Thus indeed the eigenvalues of $\Pk$ are zero and $1$, making it a projection operator obeying
\be
P_{\omega^{p'},{k}}\, \Pk = \delta_{pp'} \Pk\ .
\label{PPP}
\ee
Projectors with different $k$ simply commute. This is essential to having the energy levels be independent of how they are filled (i.e.\ which values of $p_k$ are chosen). Thus these considerations give a more formal way of characterizing what it means to be a free parafermion theory: there exist 
shift operators and {\em commuting} projectors obeying (\ref{PssP}), (\ref{PPP})
and (\ref{Hpsik}).

\subsection{Generalizing the Clifford algebra}
\label{sec:algebra}

The spectrum of all the Hamiltonians was computed by a rather elaborate sequence of computations. This was necessary because it was not obvious how to derive a useful generalization of the Clifford algebra (\ref{Cliffk}) for arbitrary $n$. I say ``useful'' in the sense of ``useful for computing the spectrum''.
\comment{ The parafermions defined in (\ref{psidef}) do provide a very interesting generalization of the Clifford algebra of the Majorana fermions, but as described in section \ref{sec:shift} in general how to use it to find the spectrum.}
If one could derive directly the algebra of the shift operators, then the spectrum could be found without the detailed analysis of sections \ref{sec:Hwn} and \ref{sec:spectrum}. In fact, just one relation is needed, having $(\Psi_{+k}+\Psi_{-k})^2\propto 1$ for the fermions generalize to
\be 
\Big(\sum_{p=0}^{n-1}\Psi_{\omega^p,k} \Big)^n \propto 1\ .
\label{sumPn}
\ee
This shows no state is annihilated by all the $\Psi_{\omega^p,k}$, and so the only way the commutation relation (\ref{Hshift}) can be consistent for all $nL$ shift operators is for the spectrum to be given by (\ref{spectrum}). 
Despite the explicit expressions for the shift operators, trying compute (\ref{sumPn}) by brute force is a nightmare, hence the necessity of the elaborate sequence of computations involving the higher Hamiltonians.  Put another way, there is no obvious way to do a Bogoliubov transformation in the same fashion as for fermions done in section \ref{sec:bogo}, because there is no obvious way of rewriting the individual terms in the Hamiltonian in terms of the shift operators.

So it goes. But now that the hard work in finding (\ref{higherspectrum}) has been done, it is possible to {\em infer} some of the algebra generalizing (\ref{Cliffk}). 
Some relations among the commutators are necessary given the spectrum. Shifting with the same operator twice would result in an energy not in (\ref{higherspectrum}). Thus
$$\Psi_{\omega^p,k}^2=0\ .
$$
In fact, for the same reason, more generally
\be \Psi_{\omega^p,k}^{}\Psi_{\omega^{p'},k}=0\ ,\quad\hbox{for }p'\ne p+1\ .
\label{sspp}
\ee
This does {\em not} vanish when $p'=p+1$; this is a sequence of clockwise shifts. In addition, shifting the $k$th level and then the $k'$th energy levels has the same effect no matter which order the shifts are done. One might be tempted to conclude  $\Psi_{\omega^p,k}^{}$ and $\Psi_{\omega^{p'},k'}$  must commute for $k\ne k'$, but that is not necessary; they need only give the same state up to some constant. Thus
\be
 \Psi_{\omega^p,k}^{}\Psi_{\omega^{p'},k'} \propto \Psi_{\omega^{p'},k'} \Psi_{\omega^p,k}^{}\ \quad\hbox{for }k'\ne k\ .
 \ee
Explicit computation confirms that indeed they do not commute, and also suggests both the value of the constant and a relation valid for all $k,k,p,p'$:
 \be
(\omega^p\epsilon_k-\omega^{p'-1}\epsilon_{k'}) \Psi^{}_{\omega^p,k}\Psi_{\omega^{p'},k'} =(\omega^{p'}\epsilon_{k'}-\omega^{p-1}\epsilon_{k})\Psi_{\omega^{p'},k'} \Psi^{}_{\omega^p,k}\ . 
 \label{ssconj}
 \ee
It would very interesting to prove this.

The algebra of the projection operators (\ref{Pdef}) was worked out in (\ref{Pcomm}) and (\ref{PPP}). Their algebraic relations with the shift operators are given in (\ref{PssP}). One additional interesting relation also holds. Because shifting clockwise $n$ times must give something proportional to the original state, 
\be
P_{\omega^p,k} \propto \Psi_{\omega^p,k}\Psi_{\omega^{p+1},k} \dots \Psi_{\omega^{p+n-1},k}\ .
\label{Psss}
\ee
This indeed commutes with the Hamiltonian and annihilates any state other than that with contribution $\omega^{p}\epsilon_k$ to the energy. Moreover, this and (\ref{Pcomm}) are consistent with the conjecture (\ref{ssconj}). 

A similar argument leads to (\ref{sumPn}). Due to (\ref{sspp}), most of the terms vanish when multiplying out the product. The only ones remaining are those of the form (\ref{Psss}). Since the $\sum_{p=0}^{n-1}\Pk=1$, this indeed yields (\ref{sumPn}), and shows that the unknown normalization in (\ref{Psss}) is independent of $p$. It is worth recalling here that as noted at the end of section \ref{sec:vn}, computer evidence suggests that the (sort-of) analogs of the Majorana fermions obey
$$(\vv{m})^n\propto 1\ .$$
This implies (\ref{sumPn}), and the converse seems a very plausible, if not yet proven, result. 

The existence of the analogs of the Majorana fermions with at least some nice algebraic properties hints
that there is more algebraic structure to be uncovered. Since all the higher Hamiltonians for $n=2$ are bilinears in the fermions, it is natural to expect that there also exist analogs of the bilinears with nice algebraic properties for general $n$. It indeed is possible to to rewrite the Hamiltonians in terms of operators arising in the same fashion as  $\vv{m}$ arose. Just like $\HH^m\psi_1$ for $m\ge nL$ can be written in terms of those with $m<nL$, it follows from (\ref{HX}) that the full Hamiltonians have a similar property. Defining the coefficients of the polynomial 
$Q_{2L}(u)$ from (\ref{Qdef}) via
$$Q_{a}(u) = u^{(1-(-1)^a)/4}\sum_{j=0}^L{\cal Q}_{a,j} u^{j}$$ 
so that
$$
\sum_{j=0}^{L} {\cal Q}_{2L,j}\hhh{jn+q} = \sum_{k=1}^{2L} Q_{2L}(u_k)\Xk = 0
\label{sumcalQ}
$$
Splitting $\hhh{m}=\sum_{b=1}^{2L-1} h^{(m)}_b$ in the obvious way (e.g.\ $h^{(1)}_b=h_b$), and then defining 
$$\chi^{(m)}_b = \sum_{j=0}^{s}{\cal Q}_{2[m/n]+1-\delta_{q,0},\,j}\, h^{(jn+q)}_b$$
in the same fashion as (\ref{vQ}) gives analogs of fermion bilinears.  Since they satisfy $\chi^{(2L)}_b=0$ for all $b$ there are indeed order $L^2$ of them.

A hint that the algebra of $\chi^{(m)}_b$ is nice comes from a closely related model, the superintegrable chiral Potts model. Onsager's original (pre-fermion) solution of the Ising model \cite{Onsager1} exploits the fact that the pieces of the Hamiltonian obey an interesting algebra, now known as the Onsager algebra. Rewritten in terms of the fermions, it turns out that all the elements of the algebra are fermion bilinears (the commutator of two fermion bilinears always gives another bilinear).  This fact was mostly forgotten as a consequence of Kaufman's solution using fermions, but remarkably, the pieces of the Hamiltonian of the superintegrable chiral Potts model obey the same algebra 
\cite{Dolan,Gehlen}. \comment{The superintegrable chiral Potts Hamiltonian is hermitian and closely related to Baxter's:
\be
H_{SCP}= e^{i\pi/2n} H + e^{-i\pi/2n} H^\dagger\ .
\label{scp}
\ee
so }

\section{Conclusion}
\label{sec:conclusion}

I have shown that certain $\zn$ clock models naturally generalize the free-fermionic structure of the quantum Ising chain.
This has many implications, both on the physics and on the more formal sides.
An obvious future direction is to search for hermitian Hamiltonians exhibiting free-para\-ferm\-ionic behavior. This is easy to do with non-local terms, but not so obvious how to achieve with a local interaction. One strategy would be to take two copies, and search for a way of coupling them so that levels are occupied in complex-conjugate pairs.

A more straightforward first step might be to use this algebraic structure to provide more direct derivations of exact results for the integrable chiral Potts Hamiltonians, such as the superintegrable chain. Exact computations in many vaunted models such as the Lieb-Liniger one can be done by starting with free particles and then adding interactions that leave the model integrable. Obviously, the model with Hamiltonian $H^\dagger$ is solvable by the same methods; the interesting question is what of the algebraic structure survives when the Hamiltonian includes both $H$ and $H^\dagger$. There is already considerable evidence that much of it does; in addition to the computation of the parafermionic zero modes in \cite{para}, shift operators and conserved charges have been found using these techniques \cite{CF}. At minimum this should enable a direct derivation of the Bethe equations for the integrable
chiral Potts models without using functional relations.

A related direction would be to study two-dimensional classical lattice models whose transfer matrix has the same eigenvectors as the Hamiltonians here. The free energy then will be written in terms of the same levels $\epsilon_k$, just as in the classical Ising model. In fact, Baxter has already shown \cite{Baxnew} that this analysis can be generalized to the Bazhanov-Stroganov model \cite{BS,Baxrecent}, whose transfer matrix commutes with that of the integrable chiral Potts model. Thus this should provide a method to better understand the chiral Potts model.

The non-hermiticity results in negative or non-local Boltzmann weights in such a classical model. This is a common characteristic of models where the degrees of freedom are geometric objects such as self-avoiding loops. In fact, a great breakthrough in the studies of the Ising model came with the realization that it can be written as dimer model, whose partition function is given by a Pfaffian (the square root of a determinant of an antisymmetric matrix) \cite{Kasteleyn,Fisher,McCoyWu}. The partition function of a free-fermion system is given by a Pfaffian \cite{Samuel}, and this is indeed a nice way of relating Kaufman's results to the dimer ones.  Thus it would be very exciting if such an object generalizing the Pfaffian gave the partition function for free parafermion models with $\zn$ symmetry. Such a correspondence is already hinted at by results in conformal field theory. The correlators of fermionic fields in the two-dimensional Ising model in the continuum limit of its critical point is given by a Pfaffian as well. Parafermionic conformal field theories describing the continuum limit of the (non-chiral) integrable clock model are well understood \cite{FZ1,FZ2}, and an explicit formula for all the parafermionic correlators is known \cite{RR}.  The general formula has some very elegant clustering properties, and so it is reasonable to hope that there is a lattice analog of it. Analyzing correlators in these free-parafermion chains may provide a path to understanding this.

More generally, analyzing correlators in these models might be able to answer a host of interesting questions. Is there a generalization of Wick's theorem to $\zn$-invariant theories? Is there an identifiable continuum limit of these free-parafermion theories? A tantalizing speculation is that these are somehow related to chiral parts of conformal field theories; the integrable chiral Potts model has an interesting connection to a perturbed chiral conformal field theory \cite{CardyPotts}, and so a connection here would not be completely surprising.

Less ambitiously, it would also be nice to explore further the properties of the generalized Clifford algebra introduced in section \ref{sec:algebra}. In particular, it is not clear whether the form (\ref{ssconj}) is general or particular to this model, and so it would be illuminating to understand properties independent of this representation. As mentioned there, the connection to the Onsager algebra \cite{Dolan,Davies} is particularly intriguing. It would probably be useful to generalize the results here to periodic or twisted boundary conditions, because this is where the Onsager algebra arises. 

Finally, it is worth noting that even though this paper is quite formal, recent work has shown how the physics of parafermions could potentially be realized experimentally in a variety of settings 
 \cite{Clarke,Lindner,TeoKane,BarkQi,Cheng,Vaezi,Pollmann}, including possibly in 2+1 dimensions \cite{Burrello,Mong}. Since the analysis of the parafermionic zero modes in \cite{para} was an important part of this story, I am hopeful that the results contained here will be of use in this pursuit as well.


\medskip

{\em Acknowledgments:}  I am very grateful to Rodney Baxter for his explaining his clock Hamiltonian results to me (I had read his papers \cite{Baxterclock}, but thought I misunderstood because the results seemed too good to be true). This paper would not exist without his insights. I would also like to thank Eduardo Fradkin for many conversations on parafermions, Nick Read for comments on the manuscript, and J.-S. Caux for pointing out the Newton-Girard formula (with an apology on behalf of the Anglo-Saxon world for the order of names). 
My research is supported by the National Science Foundation under the grant DMR/MPS1006549.

\vfill\eject
\appendix
\section{The explicit expression of the higher Hamiltonians}
\label{app:higher}

The proof that the ansatz (\ref{ansatz}) is correct in general utilizes the recursion relation (\ref{Jrecur}) from section \ref{subsec:conserved} . 
For ease of notation, I define 
\be
D^{(0,0)}=1\ ,\qquad\frac{A_{m0}}{\beta_0}=\frac{1}{\beta_m}\ ,\qquad
A_{0m}=1\ .
\label{Am0}
\ee
so that the ansatz is
\be
H^{(m)}_{a+1}=H_{a-1}^{(m)}+  \sum_{r=0}^{m-1}\sum_{s=0}^{m-r} \frac{\beta_m}{\beta_{m-r}}
 h_a^r A_{rs} D^{(s,m-r-s)}\ .
\label{ansatz2}
\ee
Note that the $\beta_m$ always appear in a ratio, so they can be rescaled to make $\beta_1=1$. Similarly, $A_{11}$ can be scaled to $1$.

Plugging the recursion relation (\ref{Jrecur}) and the ansatz into the definition (\ref{HJ}) of the higher Hamiltonians gives an expression where all the $h_a$ are explicit:
\begin{eqnarray}
\nonumber
&&H_{a}^{(m)}+  \sum_{r=1}^{m-1}\sum_{s=0}^{m-r} \frac{\beta_m}{\beta_{m-r}}
 h_a^r A_{rs} D^{(s,m-r-s)}
\  =\ (-1)^{m+1} m (J_a^{(m)}+h_a J^{(m)}_{a-1})\\
 &&\qquad +\sum_{q=1}^{m-1}(-1)^{m-q+1}\Big(H^{(q)}_{a}
  + \sum_{r=1}^{q-1}\sum_{s=0}^{q-r} \bb{q}{q-r}
 h_a^r A_{rs} D^{(s,q-r-s)}\Big)(J^{(m-q)}_a+h_aJ_{a-1}^{(m-q-1)})\ .\qquad
\label{ugly}
\end{eqnarray}
Thus proving (\ref{ansatz}) can be done piece by piece for each power of $h_a$. As a quick check, this relation (\ref{ugly}) holds at order $h_a^m$ as a consequence of having $J_{a-1}^{(0)}=1$, while order $h_a^0$ follows from applying (\ref{HJ}) for $H_a^{(m)}$. All the powers of $h_{a-1}$ can also be made explicit by using the rewritten ansatz (\ref{ansatz2}) and the recursion relation $J^{(m-q)}_a=J^{(m-q)}_{a-1}+h_{a-1}J^{(m-q-1)}_{a-2}$.

\subsection{An explicit example}
\label{app:explicit}

To illustrate the general procedure, it is useful to first study $m=3$, where 
\begin{eqnarray*}
&&H_a^{(3)} + h_a^3+\frac{\beta_3}{\beta_2}h_a(A_{11}D^{(1,1)}+A_{12}D^{(2,0)})+\beta_3 h_a^2 A_{21}D^{(1,0)}=3 (J_a^{(3)}+h_a J^{(2)}_{a-1})\\
&&\ +\Big(H^{(2)}_{a-1}+D^{(1,1)}+D^{(2,0)} + h_a^2+\beta_{2} h_a A_{11}D^{(1,0)}\Big)(J^{(1)}_a+h_a)-(H^{(1)}_a+h_a)(J^{(2)}_a+h_aJ^{(1)}_{a-1})
\ .
\end{eqnarray*}
Comparing terms of order $h_a^2$ gives
$$\beta_3 A_{21} D^{(1,0)}=J_a^{(1)} + \omega\beta_2 A_{11}D^{(1,0)}-J^{(1)}_{a-1}\ ,$$
where the factor of $\omega$ arises from commuting $h_{a}$ through $D^{(1,0)}$.
Since $D^{(1,0)}=J^{(1)}_a-J^{(1)}_{a-1}=h_{a-1}$,
$$\beta_3 A_{21} = 1+ \omega\beta_2 A_{11}= 1+ \omega+\omega^2$$
by using (\ref{b2A}). At order $h_a$, a number of terms cancel by using (\ref{HJ}) for $H_{a-1}^{(2)}$ times $h_a$. This leaves
$$\frac{\beta_3}{\beta_2}(A_{11}D^{(1,1)}+A_{12}D^{(2,0)})=\omega D^{(1,1)}+\omega^2D^{(2,0)}+
\beta_{2} A_{11}D^{(1,0)}J^{(1)}_a-J_a^{(2)}-\omega D^{(1,0)}J_{a-1}^{(1)}+J^{(2)}_{a-1}\ .
$$
Looking at terms with $h_{a-1}^2$ gives
$$\frac{\beta_3}{\beta_{2}}A_{12}-\omega^2= \beta_2A_{11}=1+\omega$$
by using (\ref{Dmm}) and (\ref{b2A}). 
Looking at terms with $h_{a-1}$ gives
\begin{eqnarray*}
\Big(\frac{\beta_3}{\beta_{2}}A_{11}-\omega\Big)D^{(1,1)}&=&\beta_2A_{11}D^{(1,0)}J_{a-1}^{(1)}-h_{a-1}J_{a-2}^{(1)}-\omega D^{(1,0)}J^{(1)}_{a-1}
\end{eqnarray*}
However, $D^{(1,1)}=(1+\omega)h_{a-1}h_{a-2}$ follows from its definition (\ref{Ddef}) and the explicit expression (\ref{H2}) for $H^{(2)}$, so
$$\Big(\frac{\beta_3}{\beta_{2}}A_{11}-\omega\Big)(1+\omega) = 1\ .$$
Combining this with the above and using the convention $A_{11}=1$ gives
\be
A_{21}=1\ ,\qquad A_{12}=1+\omega\ ,\qquad \beta_2 =1+\omega\ , \qquad\beta_3 = {1+\omega+\omega^2}\ .
\ee

\subsection{The coefficients}
\label{app:coeff}

To find the coefficients, first consider terms of order $h_a^{m-1}$ in (\ref{ugly}) for $m>2$. This gives
$$\beta_m  h_{a}^{m-1} A_{m-1,1} D^{(1,0)} = h_a^{m-1} J_a^{(1)} -h_a^{m-2}h_aJ_{a-1}^{(1)}+ \beta_{m-1}h_a^{m-2}A_{m-2,1}D^{(1,0)}h_a J_{a-1}^{(0)} \ .$$
Using (\ref{Ddef}) to commute the last $h_a$ to the left gives
$$\beta_m  A_{m-1,1} D^{(1,0)} = J_a^{(1)} - J_{a-1}^{(1)}+\omega\beta_{m-1} A_{m-2,1}D^{(1,0)}  \ .$$
By definition, $D^{(1,0)}=h_{a-1}=J_a^{(1)}-J_{a-1}^{(1)}$. Therefore the equality holds at order $h_a^{m-1}$ if
\be
\beta_m A_{m-1,1} = 1 + \omega\beta_{m-1}A_{m-2,1}\ .
\label{hm1}
\ee
This relation holds for all $m\ge 2$, agreeing with (\ref{b2A}) for $m=3$.

To derive a recursion relation for all the $A_{rs}$, note that the power of $h_{a-1}$ in each term in (\ref{ugly}) can also be easily extracted: in $D^{(s,m-s)}$ it is already labeled in the first superscript, while rewriting $J_a^{(m)} =J_{a-1}^{(m)}+h_{a-1}J_{a-2}^{(m-1)}$ makes explicit any other $h_{a-1}$. 
So
considering the terms proportional to  $h_a^{m-s}h_{a-1}^s$ in (\ref{ugly}) gives
\begin{eqnarray*}
\frac{\beta_m}{\beta_{s}}A_{m-s,s}D^{(s,0)} = \omega^s\frac{\beta_{m-1}}{\beta_{s}} A_{m-s-1,s}D^{(s,0)}J_{a-1}^{(0)}
+ \frac{\beta_{m-1}}{\beta_{s-1}}A_{m-s,s-1}D^{(s-1,0)}h_{a-1}J^{(0)}_{a-2}\ .
\end{eqnarray*}
Using (\ref{Dmm}) and noting that this must apply for all $m=r+s$ gives
\be
\frac{\beta_{r+s}}{\beta_{s}}A_{r,s} = \omega^s\frac{\beta_{r+s-1}}{\beta_s} A_{r-1,s}
+ \frac{\beta_{r+s-1}}{\beta_{s-1}}A_{r,s-1}\ .
\label{Arsrecur}
\ee
This is a well-known recursion relation, whose solution is a {\em Gaussian binomial}. This gives 
\be
\frac{\beta_{r+s}}{\beta_s}A_{rs} = \frac{(1-\omega^{r+1})\dots (1-\omega^{r+s})}{(1-\omega)(1-\omega^2)...(1-\omega^s)}\ .
\label{betaA1}
\ee

To fix the $\beta_m$, consider the terms proportional to $h^{m-2}_{a}h_{a-1}$ for $m>2$; by connectedness these are necessarily proportional to $h_{a-2}$ as well. These give
\comment{
\begin{eqnarray}
\nonumber
\frac{\beta_m}{\beta_{t+1}} A_{m-t-1,t}D^{(t,1)}=
-\frac{\beta_{m-2}}{\beta_t}\omega^t A_{m-t-2,t}D^{(t,0)}J^{(1)}_{a-1}-
\frac{\beta_{m-2}}{\beta_{t-1}}A_{m-t-1,t-1}D^{(t-1,0)}h_{a-1}J^{(1)}_{a-2}\\
+\frac{\beta_{m-1}}{\beta_t}(A_{m-t-1,t-1}D^{(t-1,1)}h_{a-1}+A_{m-t-1,t}D^{(t,0)}J_{a-1}^{(1)})+\omega^t \frac{\beta_{m-1}}{\beta_{t+1}}A_{m-t-2,t}D^{(t,1)}
\ .
\label{3wide}
\end{eqnarray}
}
\begin{eqnarray*}
\Big(\frac{\beta_m}{\beta_2}A_{m-2,1}-\omega\frac{\beta_{m-1}}{\beta_{2}}A_{m-3,1}\Big)D^{(1,1)}&=&(\beta_{m-1}A_{m-2,1}
-\omega \beta_{m-2}A_{m-3,1})h_{a-1}J_{a-1}^{(1)}-h_{a-1}J^{(1)}_{a-2}\\
&=&h_{a-1}(J_{a-1}^{(1)}-J_{a-2}^{(1)})=h_{a-1}h_{a-2}\ ,
\end{eqnarray*}
by using (\ref{Dmm}),  (\ref{D0m}), (\ref{Am0}) and (\ref{hm1}).
Since $D^{(1,1)}=(1+\omega)h_{a-1}h_{a-2}$ and $\beta_2=(1+\omega)$, this yields
$${\beta_m}A_{m-2,1}-\omega\beta_{m-1}A_{m-3,1}=1\ .$$
It follows from (\ref{betaA1}) that 
$$\beta_{r+2}A_{r1}=\frac{\beta_{r+2}}{\beta_{r+1}}\frac{1-\omega^{r+1}}{1-\omega}\ ,$$
so  
$$\frac{\beta_m}{\beta_{m-1}}\frac{1-\omega^{m-1}}{1-\omega}-\omega\frac{\beta_{m-1}}{\beta_{m-2}}
\frac{1-\omega^{m-2}}{1-\omega}= 1\ .$$
Given that $\beta_1=1$ and $\beta_2=(1+\omega)$, this gives (\ref{betam}):
$$
\beta_m =\frac{1-\omega^m}{1-\omega}\ 
$$
for $m\ge 1$.
The quantity $\beta_0$ always cancels as a consequence of connectedness, so its value is arbitrary.
Plugging this into (\ref{betaA1}) gives (\ref{Ars}):
$$
A_{rs} = \frac{(1-\omega^{r+1})(1-\omega^{r+2})\dots (1-\omega^{r+s-1})}{(1-\omega)(1-\omega^2)...(1-\omega^{s-1})}
=  \frac{(1-\omega^{s})\dots (1-\omega^{r+s-1})}{(1-\omega)...(1-\omega^{r})}\ .
$$


\subsection{The full proof}
\label{app:proof}

A useful concept here and in the computation of the shift operators is the {\em width} of an operator. Since the higher Hamiltonians and shift operators are connected, each term has a width, defined as how many different $h_b$ appear in the operator. The maximum width for any term $\hh{m}{a}$ is the smaller of $a-1$ and $m$. The ansatz implies that terms of all widths up to the maximum appear in each $\hh{m}{a}$. 

The results of the previous subsection already give the exact formula for any width-two terms in $\hhh{m}$. These are ``translation invariant'' in the sense that the coefficients are independent of the location:
$$ \hh{m}{a}= \sum_{b=1}^{a-1} \left(h_b^m\ +\ \sum_{r=1}^{m-1}\bb{m}{r} A_{m-r,r}\, h_{b+1}^{m-r}h_b^r\  +\ \hbox{ larger-width terms}\right)\ .$$
Thus to prove that the formula (\ref{Hexp}) gives all the higher $\hhh{m}$, for a width $W$ term it sufficient to study the higher Hamiltonians $\hh{m}{W+1}$. The strategy is then to build up the width $W+1$ terms from those with width $W$ recursively by demanding that they commute with $H$.

The Hamiltonian for $a=3$ contains only width-one and width-two terms:
$$H_3^{(m)} = \sum_{s=0}^{m} \bb{m}{s}A_{m-s,s} h_2^{m-s}h_1^s \ .$$
By construction this commutes with $H_3=h_1+h_2$:
\begin{eqnarray*}
[h_1+h_2,H_3^{(m)}]&=& \sum_{s=0}^m A_{m-s,s}\bb{m}{s} \left((1-\omega^s)h_2^{m-s+1}h_1^s + (\omega^{m-s}-1)h_2^{m-s}h_1^{s+1}\right) \\
&=&\sum_{s=0}^{m-1}\left(\beta_m(1-\omega)A_{m-s-1,s+1}+ (\omega^{m-s}-1)\bb{m}{s}A_{m-s,s}\right)h_2^{m-s}h_1^{s+1}\\
&=& 0 
\end{eqnarray*}
because
$$A_{r-1,s+1}=\bb{r}{s}A_{r,s}\ .
$$
This of course does not commute with $h_3$:
$$[h_3,H_3^{(m)}]=\beta_m \sum_{s=0}^m A_{m-s,s}A_{s0} (1-\omega^{m-s})h_3h_2^{m-s}h_1^s\ .$$
To relate this to $\hh{m}{4}$, consider
$$
[H_4,\,\hh{m}{4}-\hh{m}{3}]=
[h_1+h_2+h_3,\ \hh{m}{4}-\hh{m}{3}] =  -[h_3,\, H_3^{(m)}]\ 
$$
because $H_4$ commutes with $\hh{m}{4}$. The right-hand-side contains a single $h_3$, since $\hh{m}{3}$ by definition does not. However, note that by ``translation invariance'' all terms in $\hh{m}{4}-\hh{m}{3}$ must contain $h_3$ to some non-zero power. Thus define
$X^{(r,m-r)}_a$ to be the sum of all terms in $\hh{m}{a}$ that include $h_{a-1}^r$. (In the notation used above, $D^{(r,m-r)}=X_{a-1}^{(r,m-r)}$.) This yields
\be
[h_1+h_2,X^{(1,m-1)}_4]= -[h_3,\, H_3^{(m)}]\ .
\label{h3H3}
\ee

To work out explicitly what $X^{(1,m-1)}_4$ must be, expand it in a series
$$X^{(1,m-1)}_4=\sum_{s=0}^{m-2} C_{1,m-s-1,s} h_3 h_2^{m-s-1}h_1^s\ $$
for some coefficients $C_{1,m-r,r}$. 
Note that the upper limit of the sum over $s$ is such that the exponent of $h_2$ is non-zero; otherwise this would not be connected. Then  (\ref{h3H3}) requires
\begin{eqnarray*}
[h_1+h_2,X^{(1,m-1)}_4]&=&\sum_{s=0}^{m-1}C_{1,m-s-1,s}\left( (\omega-\omega^s)h_3 h_2^{m-s}h_1^s+
 (\omega^{m-s-1}-1)h_3 h_2^{m-s-1}h_1^{s+1}\right)\\
 &=&\sum_{s=1}^{m-1}\left( C_{1,m-s-1,s}(\omega-\omega^s)  + 
 C_{1,m-s,s-1}(\omega^{m-s}-1)\right)
 h_3 h_2^{m-s}h_1^s
 \ .
 \end{eqnarray*}
Therefore
$$
C_{1,m-s-1,s}(\omega-\omega^s)  + 
 C_{1,m-s,s-1}(\omega^{m-s}-1)= -\bb{m}{s} A_{m-s,s} (1-\omega^{m-s})
\label{c1}
$$
for all $m$ and $s$. Because $C_{10s}=0$ by connectedness, this means that  
$$C_{11s-1}=\bb{s+1}{s} A_{1s}=\beta_{s+1}=\bb{s+1}{s}A_{11}\left(\bb{s}{s-1}A_{1,s-1}\right)\ .$$
The last rewriting may seem perverse, but the expression in parentheses is precisely the coefficient of the $h_2h_1^{s-1}$ term in the width-two part of $\hh{s}{3}$. Thus this case agrees with the ansatz (\ref{ansatz}). The remaining $C_{1rs}$ can be determined recursively by using
$$C_{1,r,s-1}= \omega\bb{s-1}{r} C_{1,r-1,s}+\bb{r+s}{s} A_{rs}\ .$$
To express the result in a convenient form, define $\widetilde{A}_{1,r,s}$ via
$$C_{1,r,s}= \widetilde{A}_{1,r,s} \bb{r+s+1}{s} A_{r,s}\ .$$
The recursion relation for $C$ then becomes one for $\widetilde{A}$:
$$\widetilde{A}_{1,r,s-1}\bb{s}{s-1}= \omega\bb{s-1}{r} \widetilde{A}_{1,r-1,s}A_{r-1,s}
+ A_{rs}\ .$$
Simplifying this using the explicit expressions gives
$$\widetilde{A}_{1,r,s-1}=\omega \widetilde{A}_{1,r-1,s}\bb{s-1}{s}+\bb{r+s-1}{s}
$$
Given that $\widetilde{A}_{1,0,s}=0$, this gives 
\be
\widetilde{A}_{1,r,s}=\beta_r\ =A_{1r}\ .
\ee
This is independent of $s$! This is necessary for the ansatz to be true.

The rest of the width-3 terms are determined in an analogous fashion. Defining
$$X^{(q,m-q)}_4=\sum_{s=0}^{m-q-1} C_{q,r,s} h_3^q h_2^{r}h_1^s\ $$
with $q+r+s=m$
and doing the commutators as above gives the requirement that
\be
C_{q,r-1,s}(\omega^q-\omega^s)  + 
 C_{q,r,s-1}(\omega^{r}-1) + C_{q-1,r,s} (1-\omega^{r})\ =\ 0\ .
\label{cs}
\ee
 Again exploiting the fact that $C_{q0s}=0$ by connectedness and using the explicit expression for $C_{1,r,s}$ determined above allows all the $C_{qrs}$ to be fixed uniquely recursively.
The ansatz requires that the coefficients be 
\be
C_{q,r,s}= \bb{r+s+q}{s} \,A_{qr}A_{rs}\ .
\ee
for all $q$. Plugging this into (\ref{c1}) and using the explicit expressions gives
gives
$$(1-\omega^{r-1})(\omega^q-\omega^s)-(1-\omega^s)(1-\omega^{q+r-1})+(1-\omega^{r+s-1})(1-\omega^{q}) = 0 
$$
as required.

Thus this requirement of commuting with $H$ and the fact that the width-two terms are known gives all the width-three terms. All the $\hh{m}{a}$ can be worked out by repeating this procedure using the same logic. Namely, the ``translation invariance'' determines most of the terms in $H_{a+1}^{(m)}$ of up to width $a$. Connectedness means the remaining ones are fixed {\em uniquely} by the requirement that  $[H_{a+1},\hh{m}{a}]=0$.
Thus to verify the ansatz, one merely needs to check that the resulting explicit expression (\ref{Hexp}) commutes with $H$. 

This is a straightforward exercise. Consider a particular term in (\ref{Hexp}) with a given set of exponents $r_1\dots r_W$, all non-zero by connectedness. The result of commuting this term with $H$ gives terms where one of the exponents is increased $r_l\to r_l+1$, while the others stay the same. 
The resulting exponents are denoted $s_0,s_1,s_2\dots s_{W+1}$; one of the ``end'' exponents $s_0$ or $s_{W+1}$ may be non-zero even though $r_0=r_{W+1}=0$. Thus
$$
[H,\hhh{m}]=\sum^{(m)}_{\{r_j\}}\bb{m}{r_1} \prod_{j=1}^{W}  A_{r_{j+1}r_j}
\sum_{l=0}^{W+1}(\omega^{s_{l+1}}-\omega^{s_{l-1}})
\prod_{j'=0}^{W+1} h_{j'}^{s_{j'}} 
$$
where for notation's sake, $s_{-1}=s_{W+2}=0$, and $s_{j'}=r_{j'}+\delta_{j'l}$. The subscript on the first sum emphasizes that this is still a sum over allowed $r_j$, not $s_j$. The next task is to group terms with the same set of exponents $s_{j'}$. Using the explicit form (\ref{Ars}) relates the coefficients through
$$
\frac{A_{s,s'-1}A_{s'-1,s''}}
{A_{s,s'}A_{s',s''}}= \bb{s'-1}{s+s'-1}\bb{s'}{s'+s''-1}
$$
Then grouping terms with the same $s_j$ gives 
$$
[H,\hhh{m}]=\sum^{(m)}_{\{r_j\}}\bb{m}{s_1}
\sum_{l=l_{\rm min}}^{W+1}(\omega^{s_{l+1}}-\omega^{s_{l-1}})\bb{s_l-1}{s_{l+1}+s_l-1}\bb{s_l}{s_{l}+s_{l-1}-1}
\prod_{j=1}^{W} A_{s_{j+1}s_j}
\prod_{j'=0}^{W+1} h_{j'}^{s_{j'}} 
$$
where note that $A_{s_11}=1$ for all $s_1$ so that this term need not be included when $s_0=1$. Note also that the $1/\beta_{r_1}$ results in the $l=1$ term in the sum being of this form. The value of $l_{\rm min}=0$ if $s_0=1$, and is $l_{\rm min}=1$ if $s_0=0$.

The sum over $l$ looks ugly but is not difficult to do. First consider the case where $s_0=0$, and add the  $l=1$ and $l=2$ terms:
$$
 (\omega^{s_{2}}-1)\bb{s_1}{s_{2}+s_1-1}
 + (\omega^{s_{3}}-\omega^{s_{1}})\bb{s_2-1}{s_{3}+s_2-1}\bb{s_2}{s_{2}+s_1-1} =(\omega^{s_3}-1)\bb{s_2}{s_3+s_2-1} \ .
 $$
This is independent of $s_1$ and is the same form as the $l = 1$ piece with $s_1\to s_2$  and $s_2\to s_3$.  Repeating this gives for any $W'$
\be
\sum_{l=1}^{W'+1}  (\omega^{s_{l+1}}-\omega^{s_{l-1}})\bb{s_l-1}{s_{l+1}+s_l-1}\bb{s_l}{s_{l-1}+s_l-1}
=(\omega^{s_{W'+2}}-1)\bb{s^{}_{W'+1}}{s_{W'+2}+s_{W'+1}-1}
\label{sumHH}
\ee
Since $s_{W+2}=0$ by definition of $W$, the sum for $W'=W+1$ gives zero. The sum for $s_0=1$ is done in the same way since $s_{-1}=0$, and so vanishes as well. Thus
$$[H,H^{(m)}]=0$$
as claimed, and the expression (\ref{Hexp}) holds for all higher Hamiltonians.

\section{Details of the computation of shift operators}

\subsection{Relating $\HH^m\vz$ to $[\hhh{m},\vz]$}
\label{app:shift}

By the explicit expression (\ref{Hexp}) for $\hhh{m}$, the commutator can be written explicitly as
\be
[\hhh{m},\psi]=
(1-\omega^m) \sum^{(m)}
\prod_{b=1}^{{\rm w}} A_{r_{b+1}r_{b}}h_{b}^{r_b}\psi_1\ ,
\label{Hpsi}
\ee
where as before $\sum^{(m)}$ means the sum over all $r_b\ge 1$ and ${\rm w}$ such that $\sum_{b=1}^{\rm w}{r_b}=m$. Note that as a result of this commutator, the $1/\beta_r$ in (\ref{Hexp}) is cancelled, and that the only terms contributing have $h_1$ with a non-vanshing exponent.

A recursive proof of (\ref{Hmpsi}) requires evaluating the commutator $[H,[\hhh{m},\psi]]$. Its computation can be simplified by exploiting  $[H,\hhh{m}]=0$:
\begin{eqnarray*}
[H,[\hhh{m},\psi]]=[\hhh{m},[H,\psi]]=(1-\omega)[\hhh{m},h_1\psi_1]\ 
\end{eqnarray*}
Using the explicit expression  (\ref{Hexp})  gives
$$
[\hhh{m},h_1\psi_1]
= 
\beta_m \sum^{(m)}\left(
\prod_{b=1}^{{\rm w}} \frac{1-\omega^{r_1+r_2}}{1-\omega^{r_1}}A_{r_{b+1}r_{b}}h_{b}^{r_b}h_1\psi_1\ 
+\prod_{b=1}^{{\rm w}} A_{r_{b+1}r_{b}}h_{b+1}^{r_b}h_1\psi_1\right) ,
$$
where the second term arises from terms in $\hhh{m}$ that have no $h_1$ in them (i.e.\ those with $b=2$ in (\ref{Hexp})). Thus the second term has only one power of $h_1$, while since $r_1\ge 1$ in the first term,  the power there is always at least 2. It follows from (\ref{Ars}) that
$$\frac{1-\omega^{r_1+r_2}}{1-\omega^{r_1}}A_{r_{2}r_1} = A_{r_2,r_1+1}\ .$$ 
Using this and the fact that $A_{r_21}=1$ gives
$$
[\hhh{m},h_1\psi_1]
=\beta_m \sum^{(m+1)}
\prod_{b=1}^{{\rm w}} A_{r_{b+1}r_{b}}h_{b}^{r_b}\psi_1\ $$
where here $\sum_b{r_b}=m+1$.
Up to a constant factor, this is exactly the same expression that results from (\ref{Hpsi}) for $\hhh{m+1}$! 
Namely, 
\be [H,[\hhh{m},\psi]]=(1-\omega)\frac{1-\omega^m}{1-\omega^{m+1}}\ [\hhh{m+1},\psi].
\label{hmmp1}
\ee
Since (\ref{Hmpsi}) is obviously true for $m=1$, using recursion with (\ref{hmmp1}) then implies (\ref{Hmpsi}).

In this proof, there are factors of $\omega^{r}-1$ in the denominator, but these all cancel in the end. Thus as long as $\omega$ is treated as not a root of unity when the commutators are done, and then then limit $\omega^n\to 1$ is taken, these results remain true. It is also worth noting that this simple result holds only if one starts with certain operators like $\psi_1$ (or $\psi_2$, $\psi^\dagger_{2L-1}$, $\psi^\dagger_{2L-2}$).

\subsection{The derivation of the shift operators}
\label{app:shiftproof}

Here I give an alternative construction of the shift operators. This is less intuitive than that given in section \ref{sec:vn}, but has the very important advantage that the ``truncation'' $\vv{Ln}=0$ needed there is proven. 

Most of the technicalities needed to show this have already been worked out in section \ref{sec:Hwn}. Using the relation (\ref{Hs}) with the simplified expression (\ref{HMx}) for the higher Hamiltonians yields a similar expression for $\HH^m\vz$.  A key fact is that there is a maximum value of $m$ in the reduced, fixed-width operators: $\xb{m}{\ww+b,b}=0$ for $m\ge nL$ as a consequence of the exclusion rule (\ref{exc}) combined with $\ww<2L$.  Thus the higher Hamiltonians with $m\ge nL$ are written as a sum over the same set of operators as the lower ones are. Moreover, the coefficients of all these Hamiltonians involve the same matrix ${\cal M}$.


The first step in finding the shift operators is to define reduced, fixed-width, terms analogous to $\xmw$, but with non-zero $\zn$ charge. 
All $\xmw$ with $b\ne 1$ commute with $\psi_1$, and using $[h_1^{r_1},\psi_1]=(1-\omega^{r_1})h_1^{r_1}\psi_1$ gives
\be
\umw=\lim_{\omega^n\to 1} \frac{1}{1-\omega^m} [\xmw,\psi_1] = 
\sum^{(m,\ww)}\,\left(\prod_{j=1}^\ww A_{r_{j+1}r_{j}}h_{j}^{r_j}\right)\psi_1\ ,
\label{umwdef}
\ee
with as before $\sum^{(m,\ww)}$ defined as the sum over all $1\le r_j<n$ such that $\sum_{j=1}^{\ww} r_j =m$. Note that as explained in the previous subsection, $\ub{sn}{\ww}$ is non-trivial even though $\hhh{sn}$ is proportional to the identity. Using the simplified expression (\ref{HMx}) gives
\be
\HH^m\psi_1 =  \sum_{\ww=0}^{2L-1}\ \sum_{s=0}^{[m/n]}
 ({\cal M}^{2s+\ww })_{\ww+1,1}\, \ub{m-ns}{\ww}\ ,
 \label{HMxs}
\ee
with $\ub{m}0\equiv\delta_{m0}\psi_1$.

\comment{
\begin{eqnarray*}
\HH^m\psi_1 &=& \sum_{j=1}^{L} \sum_{\ww=1}^{2L-1}\ \sum_{s=0}^{[m/n]}
 ({\cal M}^{\ww-2s })_{2j+\ww+1,2j+1}({\cal M}^{2[m/n]})_{2j+1,1}\, \ub{M+ns}{\ww}\\
 &=&\sum_{j=1}^{L} X_{2j}^{(M)}({\cal M}^{2[m/n]})_{2j+1,1}
\end{eqnarray*}
}


The shift operators are written in terms of the eigenvectors of ${\cal M}$, which is simply the matrix ${\cal F}$ used to solve the free-fermion model in a rescaled basis. Thus the eigenvectors and the corresponding orthogonality relations follow from a rescaling of polynomials $Q_a$ defined in (\ref{Qdef}). 
Since ${\cal M}$ is not symmetric like ${\cal F}$, its right and left eigenvectors are different. Rescaling (\ref{evM}) gives the left eigenvectors 
\be
\mu_{L,b}= (\pm 1)^{b}\, Q_{b-1}(u_k),\qquad \hbox{ where }
\sum_{a=1}^{2L} \mu_{L,a} {\cal M}_{ab} =\pm u_k^{1/2}\mu_{L,b} .
\label{evM}
\ee
For this to be an eigenvector, the $u_k$ must be roots of $Q_{2L}(u)$, i.e.\ $Q_{2L}(u_k)=0$.
The index $k=1\dots L$ because $Q_{2L}(u)$ is a polynomial of order $L$. 
\comment{Since ${\cal M}^2$ splits into blocks acting on even and odd indices, the eigenvectors do as well. Rescaling (\ref{evFsq}) gives
\be
 \sum_{j'=1}^L Q_{2j'-1}(u_k)({\cal M}^2)_{2j',2j}=u_k Q_{2j-1}(u_k)\ ,\qquad
\sum_{j'=1}^L Q_{2j'-2}(u_k)({\cal M}^2)_{2j'-1,2j-1}=u_k Q_{2j-2}(u_k)\ .
\label{evMsq}
\ee
}
The right eigenvectors of ${\cal M}$ are given by the rescaled version of $Q_a$ from (\ref{QQdef}):
$$\mu_{R,a}= (\pm 1)^{a}\ \QQ{a-1}(u_k)\ .
$$
In terms of the $\QQ{m}$, the orthonormality relations (\ref{ortho}) become
\be
\sum_{j=0}^{L-1} \QQ{2j}(u_k)Q_{2j}(u_{k'}) = \sum_{j=0}^{L-1} \QQ{2j+1}(u_k)Q_{2j+1}(u_{k'}) = N_k \,\delta_{k,k'}\ \ ,
\label{orthobar}
\ee
while (\ref{orthom}) becomes
\be
\sum_{k=1}^L u_k^{m/2}\frac{1}{N_k}\, \QQ{a-1}(u_k)Q_{a'-1}(u_k) =  ({\cal M}^m)_{a,a'}\ ,
\label{orthombar}
\ee
applicable when $(-1)^{m+a+a'}=1$. 

This orthogonality relation (\ref{orthombar}) and the fact that $Q_0(u)=1$ allow (\ref{HMxs}) to be rewritten as
\begin{eqnarray}
\nonumber
\HH^m\psi_1 &=& \sum_{k=1}^L  \sum_{\ww=0}^{2L-1} \sum_{s=0}^{[m/n]}
\frac{1}{N_k} u_k^{[m/n]-s+\ww/2} \QQ{\ww}(u_k)\,\ub{sn+q}{\ww}\\
&=&\sum_{k=1}^L  u_k^{[m/n]}\, \Phi^{(q)}_k
\label{HmPhi}
\end{eqnarray}
where $q=m\,$mod$\,n$, $s$ has been redefined by $s\to [m/n]-s$,  and
\be
\Phi^{(q)}_k =\sum_{\ww=0}^{2L-1}\sum_{s=0}^{s_{\rm max}} \frac{1}{N_k}u^{-s+{\ww/2}} \QQ{\ww}(u_k) \ub{sn+q}{\ww}
\label{phiep1}
\ee
where $s_{\max}$ is the maximum value of $s$ such that $\ub{sn+q}{\ww}$ is non-vanishing, i.e.\
$s_{max}n+q\le n[(\ww+1)/2]+((-1)^\ww-1)/2$. This follows from the exclusion rule (\ref{exc}), and is explained following (\ref{xmw}). The precise expression is not particularly important; what is important is that $s_{\rm max}$ does not grow with $m$, so the only dependence on $m$ on the right-hand-side of (\ref{HmPhi}) is in the $u_k^{[m/n]}$, and in $q$. 

It is easy now to show that $\HH^m\psi_1$ for $m\ge nL$ is given by a linear combination of those with $m<nL$. Since $Q_{2L}(u)$ defined in (\ref{Qdef}) is a polynomial in $u$, $Q_{2L}(\HH^n)\psi_1$ is a linear combination of commutators. This particular linear combination vanishes:
\be 
Q_{2L}(\HH^n)\psi_1= \sum_{k=1}^{L} Q_{2L}(u_k)\,\Phi^{(0)}_k = 0\ .
\label{QHzero}
\ee
using (\ref{Qroots}) and  (\ref{HmPhi}) with the fact that $\Phi^{(0)}$ is independent of $m$. Since $Q_{2L}(u)=u^{L}+\dots$, this indeed gives $\HH^{nL}$ in terms of lower ones.

Thus the space of operators obtained by commuting $\psi_1$ with the $\hhh{m}$ is finite-dimensional when $L$ is finite, whereas (\ref{HmPhi})  holds for all positive integers $m$. Thus the only way for this to be consistent is for the $\Phi^{(q)}(u)$ to be eigenvectors of ${\cal H}^n$:
\be
\HH^n\Phi^{(q)}_k = u_k \Phi^{(q)}_k\ .
\label{evHn}
\ee
This also can be (and has been) verified by explicitly working out the commutator $\HH\xmw$. Much easier is to exploit (\ref{QHzero}) to find a nice basis; this is done in section \ref{sec:vn}.

As a consistency check, note that the $\HH^m\vz$ can be written in terms of the shift operators as
\be
\HH^m\vz = \frac{1}{n}\sum_{k=1}^{2L}\sum_{p=0}^{n-1}(\epsilon_k\omega^p)^m \,\Psi_{\omega^p,k}\ .
\label{Hshift}
\ee
This can be proved directly using the $\vv{m}$. The only non-trivial part is to prove the expression for $m=0$; the result for all $m$ immediately follows 
because $\Psi_{\omega^p,k}$ is an eigenstate of $\HH$ with eigenvalue $\omega^p\epsilon_k$. The right-hand side of (\ref{Hshift}) for $m=0$ is indeed
\begin{eqnarray*}
\frac{1}{n}\sum_{k=1}^{2L}\sum_{p=0}^{n-1} \Psi_{\omega^p,k} 
=
\frac{1}{n}\sum_{k=1}^{2L} \sum_{p=0}^{n-1}
\sum_{q=0}^{n-1} \left(\omega^{p} \epsilon_k\right)^{-q} \Phi^{(q)}_k
=\sum_{k=1}^{2L}\Phi^{(0)}_k
=\sum_{m'=0}^{L-1} \sum_{k=1}^{2L}\frac{1}{N_k}   \QQ{2m'}(\epsilon_k^n)\, \vv{m'n}\ 
= \vz
\end{eqnarray*}
by using (\ref{Psired}), (\ref{chidef}) and finally the orthogonality relation (\ref{orthombar}) for $j=0$.

\vfill\eject

\end{document}